\documentclass[journal]{IEEEtran}

\usepackage{cite}
\usepackage[cmex10]{amsmath}
\DeclareMathOperator{\rank}{rank}
\DeclareMathOperator{\Tr}{Tr}
\DeclareMathOperator{\diag}{diag}
\DeclareMathSizes{10}{9}{7}{6}
\newcommand{\vect}{\operatorname{vec}}
\newcommand{\crb}{\operatorname{CRB}}
\newcommand{\rect}{\operatorname{rect}}
\usepackage{url}

\usepackage{amssymb}   
\usepackage{amsxtra}
\usepackage{amscd}
\usepackage{amsthm}
\usepackage{amsxtra}
\usepackage{amscd}
\usepackage{amsthm}
\usepackage{tikz}
\usetikzlibrary{matrix}
\usepackage{mathtools}
\usepackage{graphicx}   
\usepackage{wrapfig}
\usepackage{layouts}
\usepackage{booktabs,makecell,multirow,tabularx}
\usepackage{arydshln}
\usepackage{flushend}
\usepackage{xparse}
\usepackage[caption=false]{subfig}
\NewDocumentCommand{\overarrow}{O{=} O{\uparrow} m}{%
	\overset{\makebox[0pt]{\begin{tabular}{@{}c@{}}#3\\[0pt]\ensuremath{#2}\end{tabular}}}{#1}
}
\NewDocumentCommand{\underarrow}{O{=} O{\downarrow} m}{%
	\underset{\makebox[0pt]{\begin{tabular}{@{}c@{}}\ensuremath{#2}\\[0pt]#3\end{tabular}}}{#1}
}

\usepackage{verbatim}
\usepackage{stfloats}
\usepackage{blindtext}
\usepackage{abraces}
\usepackage{algpseudocode}
\usepackage{algorithm}
\usepackage{color}

\newcommand\scalemath[2]{\scalebox{#1}{\mbox{\ensuremath{\displaystyle #2}}}}

\usepackage{xparse}
\renewcommand{\thefootnote}{\arabic{footnote}}
\let\svthefootnote\thefootnote

\usepackage{balance}

\newcommand*\mystrut[1]{\vrule width0pt height0pt depth#1\relax}

\usetikzlibrary{decorations.pathreplacing,
	matrix,
	positioning,
}

\usepackage[bottom]{footmisc}
\usepackage{tikz}

\usetikzlibrary{decorations.pathreplacing}
\newcommand{\vast}{\bBigg@{4}}
\newcommand{\Vast}{\bBigg@{5}}
\usepackage{float}
\usepackage{xcolor,cite,etoolbox}
\usepackage[numbers,sort&compress]{natbib}

\usepackage{filecontents}

\bstctlcite{IEEEexample:BSTcontrol}
\begin{document}
	%
	\title{Hybrid STAR-RIS Enabled Integrated Sensing and Communication}
	
	%
	%
	%
	\author{\mbox{Zehra~Yigit,~\IEEEmembership{Member,~IEEE,}
			Ertugrul~Basar,~\IEEEmembership{Fellow,~IEEE}
			\thanks{Z. Yigit  is  with  Turkcell 6GEN Lab., Istanbul, Turkiye.  E-mail: {zehra.yigit@turkcell.com.tr}.}}
		\thanks{E. Basar is with the Department of Electrical Engineering, Tampere University, 33720 Tampere, Finland, on leave from the Department of Electrical and Electronics Engineering, Koc University, 34450 Sariyer, Istanbul, Turkiye. E-mail: ertugrul.basar@tuni.fi and ebasar@ku.edu.tr.}
	}
	
	
	\maketitle
	
	\begin{abstract}
		Integrated sensing and communication (ISAC) is recognized as one of the key enabling technologies for sixth-generation (6G) wireless communication networks, facilitating diverse emerging applications and services in an energy and cost-efficient manner. This paper proposes a multi-user multi-target ISAC system to enable full-space coverage for communication and sensing tasks. The proposed system employs a hybrid simultaneous transmission and reflection reconfigurable intelligent surface (STAR-RIS) comprising active transmissive and passive reflective elements. In the proposed scheme,  the passive reflective elements support communication and sensing links for local communication users and sensing targets situated within the same physical region as the base station (BS), while low-power active transmissive elements are deployed to improve sensing performance and overcome high path attenuation due to multi-hop transmission for distant communication users and  sensing targets situated far from of the coverage area of the BS. Moreover, to optimize the transmissive/reflective coefficients of the hybrid STAR-RIS, a semi-definite relaxation (SDR)-based algorithm is proposed. Furthermore, to evaluate communication and sensing performance, signal-to-interference-noise ratio (SINR) and   Cramer-Rao bound (CRB) metrics have been derived and investigated via conducting extensive computer simulations.
		
	\end{abstract}
	

	\let\thefootnote\relax\footnote{This work was supported by The Scientific and Technological Research Council of Turkiye (TUBITAK) through the 1515 Frontier Research and Development Laboratories Support Program under Project 5229901 - 6GEN. Lab: 6G and Artificial Intelligence Laboratory, and also Grant 120E401.}
	
	\IEEEpeerreviewmaketitle
	\vspace{-0.cm}
	\section{Introduction}
	Sixth-generation (6G) networks are envisioned to provide intelligent, secure, and ubiquitous wireless connectivity via new and enhanced usage scenarios with advanced capabilities \cite{itu}. In this vision, integrated sensing and communication (ISAC) stands out as a key enabling technology that offers a unified platform for hardware resources, spectrum, and signal processing framework to enable simultaneous communication and sensing functionalities \cite{itu, liu2022integrated}. This integration brings mutual benefits of communication and sensing via enhancing the hardware, spectrum, energy, and cost efficiencies while defining new sensing-related environmental-aware applications and services, including vehicle-to-everything (V2X), remote sensing, smart cities, smart healthcare, and environmental monitoring \cite{cui2021integrating}.
	
		Although ISAC has gained significant attention in the literature over the past few years, it has a longstanding history of research under various names, such as joint radar and communication (JRC) \cite{liu2020joint}, joint communication and radar (JCR) \cite{zhang2021overview} and dual function radar communication (DFRC) \cite{hassanien2016signaling}. The primary difference between JRC and JCR lies in their respective emphasis: JRC prioritizes radar-centric designs with integrated communication capabilities, while JCR  emphasizes communication-centric designs enhanced with radar functionalities. On the other hand, DFRC refers to a unified platform that utilizes shared waveforms for both communication and sensing tasks \cite{zhang2021overview}. 
	
	Reconfigurable intelligent surface (RIS){-empowered communication}  is another emerging technology that offers  energy and cost-efficient solutions for ubiquitous  connectivity in next generation networks \cite{basar2019wireless, basar2023reconfigurable}. RISs are planar metasurfaces consisting a large number of passive reflective elements  to efficiently modify  wireless propagation environment, thereby paving the way of achieving the vision for a  smart radio environment \cite{di2020smart, wu2024intelligent}.  By dynamically shaping and managing the propagation of electromagnetic waves, RISs are primarily utilized to improve signal quality \cite{huang2019reconfigurable, basar2019transmission}, enhance channel capacity \cite{zhang2020capacity}, mitigate {inter-}user interference  \cite{wu2019intelligent}  and  ensure physical layer security \cite{chen2019intelligent}. 
	
	
	Despite their potential, RIS-assisted systems  inherently experience multiplicative path attenuation and passive RIS, equipped with  reflective  elements that can solely manipulate the phases of the incident signals, struggles to overcome this hurdle   \cite{basar2021present, ellingson2021path}. 
	To address this challenge, the concept of active RIS, capable of simultaneously controlling the magnitudes and phases of incident signals, has been proposed \cite{basar2021present,zhang2022active,zhi2022active, yigit2022hybrid}. This approach involves integrating active RIS elements with low-power reflective-type amplifiers, such as tunneling reflection amplifiers \cite{long2021active} and amplifying reflect arrays \cite{kishor2011amplifying}. While this integration allows for the adjustment of the magnitudes of incident waves, it comes with the trade-off of an increase in  power consumption and the unavoidable introduction of thermal noise.  Specifically, the reflection coefficient of an active RIS element, $\Omega_a$, and a passive RIS element, $\Omega_p$, can be defined as 
	\begin{align}
		& 	\Omega_a=\xi_{a} e^{j\phi_{a}} \:\:\text{for} \:\: 	  \xi_{a} > 1,  \:\:\text{and}\:\: \phi_{a}\in\left[-\pi, \pi \right] \label{eq1} \\
		&	\Omega_p= \xi_{p}e^{j\phi_{p}} \:\:\text{for} \:\: 	 \xi_{p} \leq 1,  \:\:\text{and}\:\: \phi_{p}\in\left[-\pi, \pi \right] \label{eq2}
	\end{align}
	where	 $	\left|  \xi_{i}\right| $ and $ \phi_{i}$ respectively are the magnitude and phase of the corresponding  RIS element for $i\in\left\lbrace a, p \right\rbrace $.

	On the other hand, unlike aforementioned  RIS architectures, which offer half-space ($180^\circ$) coverage and constrain the transmitter and receiver(s) to be located on the same side of the RIS \cite{wu2019intelligent},  simultaneous transmission and reflection RIS (STAR-RIS)  concept has been proposed to extend the RIS coverage \cite{mu2021simultaneously}.  This is achieved by enabling simultaneous reflection and transmission of the incident signal,  offering  further flexibility in the deployment of transmitter and receiver(s) \cite{mu2021simultaneously}.  In STAR-RIS, each of the STAR-RIS element splits the incoming signal from BS  in  two parts: one part of the signal is reflected on the same side of the incident signal (reflection side), while  the second part is transmitted to the opposite side of the incident signal (transmission side) \cite{mu2021simultaneously}. Therefore,   the transmission and reflection coefficients of the  STAR-RIS can be modeled as $\Omega_{t}= \xi_{t}e^{j\phi_{t}}$ and $\Omega_{r}= \xi_{r}e^{j\phi_{r}}$, respectively.   It is worth noting that controlling transmission and reflection using either a shared or separate reflection-type amplifier in each STAR-RIS element results in either coupled or independent transmission and reflection coefficients  \cite{xu2023active}.
	\subsection{Prior Works}
	Recent studies  indicate a growing  interest in deploying RIS for sensing applications \cite{elbir2022rise, liu2023integrated, chepuri2023integrated} that mostly focus on   sensing performance \cite{song2023intelligent} and  beamforming optimization \cite{xing2023joint, sankar2023beamforming} of RIS-assisted  single-target ISAC systems.\hspace{0.2cm}
	Although traditional reflection-only passive RISs can  improve  communication and sensing performance of ISAC systems in challenging non-line-of-sight (NLOS) conditions \cite{liu2023integrated, chepuri2023integrated}, they struggle to mitigate multiplicative path attenuation due to multi-hop transmission of the sensing link. This limitation   restricts their deployment to locations near the transmitter and/or receiver to achieve noticeable performance gains \cite{zhang2022active}. To overcome this limitation and enhance  performance of sensing link,  active RIS-assisted ISAC systems, which incorporates additional power amplifiers to modify not only the phase but also the magnitude of an incident electromagnetic wave, has been proposed \cite{yu2023active,zhu2023joint}.
	
	More recently, STAR-RIS-assisted ISAC systems have been introduced to achieve full-space coverage and enable flexible deployment of base-stations (BSs), sensing targets, and communication users, independent of their positioning relative to the RIS \cite{wang2023stars, zhang2023stars,  li2024star, zhang2024intelligent}. In \cite{wang2023stars},   a single-target, multi-user ISAC system is proposed,   utilizing a STAR-RIS  equipping a portion of the surface with one-side sensing-capable elements dedicated to serving sensing targets, while passive transmissive elements are employed to cater to communication users. In \cite{li2024star}, a STAR-RIS is placed on the exterior of the vehicle to improve communication for user   inside the vehicle while also enabling the vehicle to be tracked and sensed by nearby roadside units (RSUs). A multi-target multi-user ISAC scenario deploying a STAR-RIS with bi-directional sensing elements is presented in \cite{zhang2023stars}.  
	In addition, in \cite{zhang2024intelligent}, receive antennas are positioned on both sides of the STAR-RIS to mitigate severe path attenuation, enabling multi-target multi-user ISAC transmissions on both sides of the STAR-RIS. 
	
\subsection{Motivation and Contributions}
	
	Motivated by the aforementioned  studies, to further improve communication and sensing performance, as well as   energy efficiency of RIS-assisted ISAC systems while achieving  full-space coverage,  this paper proposes a novel multi-target multi-user ISAC transmission system, utilizing a dual-sided STAR-RIS, with each side offering both reflective and transmissive capabilities. In the proposed scheme, to address severe signal power degradation due to multi-hop transmission,  a hybrid STAR-RIS with active transmissive and passive reflective elements is utilized. Unlike  \cite{wang2023stars, zhang2023stars} that employ sensing capable elements and  additional receive antennas \cite{zhang2024intelligent}, leading  hardware and implementation complexity,  the proposed scheme enables an efficient  multi-target multi-user transmission on both sides of STAR-RIS with  a more cost-effective and power efficient manner. Moreover, to optimize the  transmissive/reflective coefficients of the hybrid STAR-RIS, a non-convex optimization problem is formulated, which is solved via a semi-definite relaxation (SDR)-based approach  implemented   through  CVX toolbox \cite{cvx}. Furthermore, comprehensive simulations are conducted to evaluate   signal-to-interference-noise ratio (SINR) of the communication users and sensing targets and  Cramer-Rao bound (CRB) to estimate two-dimensional (2D) angles of departure (AoDs) of the sensing targets. The main contribution of the paper can be summarized as follows:\begin{itemize}
			\item 	 In this paper, a hybrid STAR-RIS-enabled ISAC system is introduced to  enhance both communication and sensing performance. Unlike  \cite{wang2023stars} and \cite{li2024star} that partition the entire space into distinct sensing and communication regions, the proposed system enables simultaneous sensing and communication in both sides of the STAR-RIS.
			\item  To address the significant path attenuation experienced during multi-hop echo sensing signal transmission in RIS-assisted ISAC systems, the proposed system employs a novel approach compared to those presented in  \cite{xu2023active, rihan2023passive, wang2023stars, zhang2023stars, zhang2024intelligent}, which either use fully active RIS elements \cite{xu2023active, rihan2023passive}, allocate a portion of the RIS for sensing-capable elements \cite{wang2023stars, zhang2023stars} or leverage a bistatic radar setup with additional receive antennas on both sides of the STAR-RIS \cite{zhang2024intelligent}. Instead, our system utilizes a dual-sided hybrid STAR-RIS model that incorporates both passive reflective and active transmissive elements, with active transmissive elements amplifying the attenuated signal  across multi-hop transmission paths.
			\item We derive communication and sensing  SINR of the proposed scheme. Subsequently, we formulate an optimization problem that maximizes the minimum target SINR by jointly optimizing the reflection and transmission coefficients of the hybrid STAR-RIS under communication SINR constraints. Although enabling multi-user, multi-target transmission on both sides of the STAR-RIS via incorporating both passive reflective and active transmissive elements poses substantial challenges, we reformulate a non-convex optimization problem as a quadratically constrained quadratic programming (QCQP) problem. Then, this non-convex problem is  efficiently solved using a semi-definite relaxation (SDR) solution. Moreover,  we derive the CRB for estimating the 2D angles of departure (AoDs) of the targets.
			\item Thorough performance analyses are conducted via computer simulations to investigate the sensing and communication performance, as well as the power consumption of the proposed system.  Moreover, the effectiveness of the hybrid STAR-RIS-aided ISAC system is demonstrated in comparison to benchmark passive STAR-RIS-aided ISAC system.
	\end{itemize}
	
	\subsection{Organization}
	The rest of the paper is organized as follows. In Section II, the proposed system model is introduced. Section III provides derivations for sensing performance metrics  including target SINR and CRB for 2D AoDs estimation, followed by the  proposed   optimization algorithm.   In Section IV,  the numerical results are presented, and this paper  concludes with Section V.
	
	\textit{Notation:} Throughout this paper,  vectors and matrices are donated by bold-face lower letter and bold-face upper letter, respectively. $\mathbb{C}^{K\times L}$ represents the space of  a complex matrix with dimensions $K\times L$.  $|x|$ stands for absolute value of scalar $x$, while  $\left\| \mathbf{X} \right\| $ is the Frobenious norm of matrix $\mathbf{X}$.  $\mathbf{X}^{-1}$, $\mathbf{X}^\mathrm{T}$ and $\mathbf{X}^\mathrm{H}$ represent inverse, transposition and  Hermitian of the matrix $\mathbf{X}$, respectively.  For $\mathbf{x}$ being a vector, $\diag(\mathbf{\mathbf{x}})$ stands for  a diagonal matrix whose diagonal elements are equal to elements of  the vector $\mathbf{x}$. $\vect({\mathbf{X}})$ is vectorizing operation, $\rect(\cdot)$ is rectangular function, $\Tr(\mathbf{X})$ is trace value, and $\odot$ is Hadamard product. $\mathbf{I}$ denotes identity matrix, while $\mathbf{1}$ stands for a vector with all-ones elements. $\mathfrak{Re}(\cdot)$ and $\mathfrak{Im}(\cdot)$ represents real and imaginary components of a complex number, respectively. $\mathcal{CN}(\mu, \sigma^2)$ denotes distribution of a complex Gaussian random variable with mean $\mu$ and variance $\sigma^2$, and $\mathcal{O}$ denotes big-O notation.

	\begin{figure}[!t]
		\centering
		\includegraphics[width=1\columnwidth]{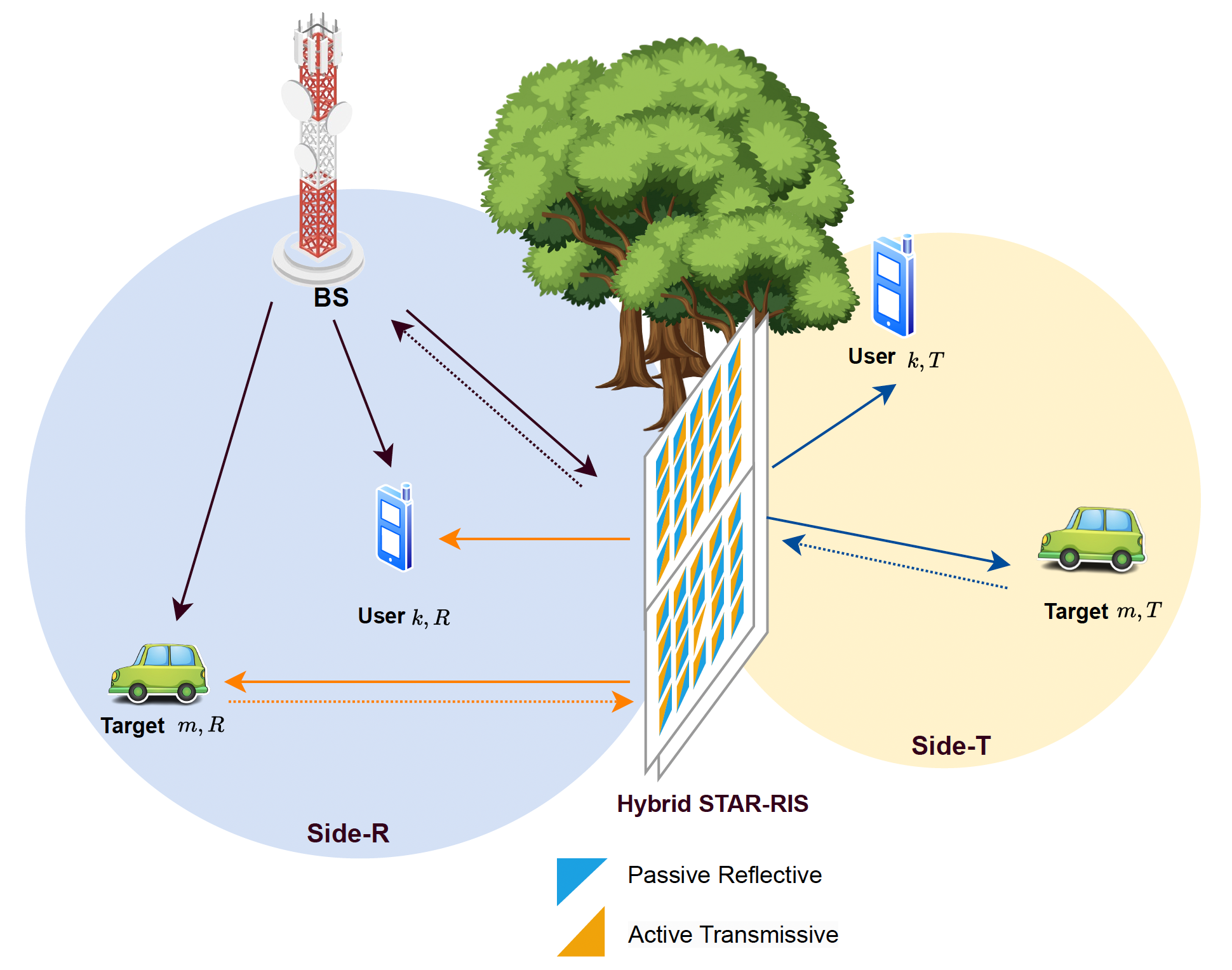}
		\vspace{-0.1cm}\caption{System model of hybrid RIS-assisted ISAC.}
		\label{sys}
	\end{figure}
	\section{System Model}
	
	In this section,   system model for the proposed hybrid STAR-RIS-enabled ISAC is presented.

	\subsection{Hybrid STAR-RIS-Enabled  ISAC}
	The system model of the  proposed hybrid STAR-RIS-assisted ISAC is illustrated in Fig. \ref{sys}.  In the proposed system,  a uniform linear array (ULA) BS with $T_x$ antennas communicates with $K$ single-antenna users, while simultaneously  performing $M$ point targets detection.{\addtocounter{footnote}{-1}\let\thefootnote\svthefootnote\footnote[1]{It is assumed that the transceiver operates in full-duplex mode to suppress self interference between joint signal transmission and echo signal reception \cite{liu2023joint}.}}  It is assumed that the coverage space of the BS is split into two parts: transmission (Side-$T$) and  reflection (Side-$R$) by a hybrid STAR-RIS with a uniform planar array (UPA) consisting  $N$ transmissive/reflective elements.  To enable two-side target detection and alleviate path attenuation, the hybrid STAR-RIS operates as a dual-sided STAR-RIS, with “STAR'' functionality present on both sides \cite{xu2022simultaneously, xu2023active}, incorporating passive reflective and active transmissive elements. Alternatively, it can be constructed using two adjacent conventional STAR-RIS units oriented oppositely, each featuring active transmissive and passive reflective elements.  However, in this paper, for the simplicity of presentation, a dual-sided STAR-RIS with unit-gain passive reflective ($\xi_{r,n}=1$) and active transmissive elements ($\xi_{t,n}>1$) is considered, where  $n\in\left\lbrace 1,2,\cdots, N\right\rbrace $. Please note that, in the proposed scheme, to independently adjust transmissive and reflection coefficients of the proposed hybrid STAR-RIS, it is assumed that two separate reflection type amplifiers are  employed. Then, the amplifiers are assumed to be configured so that on both sides, reflection coefficients maintain unity power ($\xi_{r,n}=1$)  while the transmission amplifier provides an amplification effect ($\xi_{t,n}>1$) \cite{xu2023active}. As a result, the reflection component operates in a passive mode, whereas the transmission component functions in an active mode.
	
 To better illustrate, Fig. 2 presents the signal behaviour of the hybrid STAR-RIS. An incident signal on Side-$R$, denoted as $x$, interacts with the $n$-th hybrid element, reflecting with a phase shift $e^{j\phi_{r,n}}$ as $x \times e^{j\phi_{r,n}}$ while simultaneously being transmitted toward Side-$T$ through an active transmissive element, resulting in $x' = x \times \xi_{t,n} e^{j\phi_{t,n}}$. Similarly, the arriving signal on Side-$T$, represented as $x'$, is reflected as $x' \times e^{j\phi_{r,n}}$ and transmitted back to Side-$R$ through the active transmissive element, yielding $x' \times \xi_{t,n} e^{j\phi_{t,n}}$.
		\begin{figure}[!t]
		\centering
	{	\includegraphics[width=1\columnwidth]{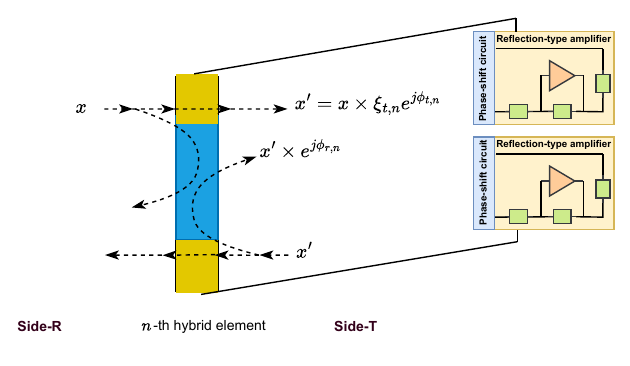}
		\vspace*{-0.6cm}	\caption{A generic signal model of the  hybrid STAR-RIS.}}
		\label{fig.sys1}
	\end{figure}

	Consequently, the transmission and reflection coefficients of the hybrid STAR-RIS with $N$ elements in both directions can be defined as:  
	\begin{align}
		&\hspace{-0.2cm} \mathbf{\Phi}_t\in\mathbb{C}^{N\times N} = \diag\left(   \xi_{t,1} e^{j\phi_{t,1}},   \xi_{t,2}e^{j\phi_{t,2}},  \cdots,  \xi_{t,N} e^{j\phi_{t,N}}\right)  \\
		&\hspace{-0.2cm} \mathbf{\Phi}_r\in\mathbb{C}^{N\times N} = \diag\left(    e^{j\phi_{r,1}},  e^{j\phi_{r,2}},  \cdots,   e^{j\phi_{r,N}}\right)  
	\end{align}
	where $\mathbf{\Phi}_t$ and $\mathbf{\Phi}_r$ are the transmission and reflection matrices, respectively.  
	
	In the proposed scheme,  it is assumed that $K_R$ users and $M_R$  targets at Side-$R$   receive both direct signals from the BS and indirect signals reflected by the hybrid STAR-RIS, while  $K_T$ users and $M_T$  targets at Side-$T$ {experience obstructed direct links from the BS and} can only perceive  transmitted signals through the hybrid STAR-RIS, where $K=K_R+K_T$ and $M=M_R+M_T$.  It is important to note that throughout the paper, regardless of their distances from the BS,  the communication users and sensing targets on Side-$R$ who are located within the same region as the BS are classified as “local", whereas those on Side-$T$ are referred as “distant".  

	In the proposed scheme, to enable simultaneous communication and sensing, the BS transmits following joint signal at $l$th time instance, represented by $\mathbf{x}(l)\in\mathbb{C}^{T_x\times 1}$ becomes
	\begin{align}
		&	\mathbf{x}(l)=\sum_{p=1}^{K}\mathbf{w}_{c,p}{c}_p(l)+\sum_{q=1}^{M}\mathbf{w}_{s,q}{s}_q(l)\label{beam1}\\
		&\hspace{0.6cm}=\mathbf{W\tilde{x}}(l)
		\label{beam}
	\end{align}
	where  $\mathbf{W}\in\mathbb{C}^{T_x\times({K+M})}$ is the overall beamforming matrix. Here, $c_p(l)$ and $s_q(l)$ are transmit signals for $p$th communication user and $q$th sensing target, while  $\mathbf{w}_{c,p}\in\mathbb{C}^{T_x\times 1}$ and $\mathbf{w}_{s,q}\in\mathbb{C}^{T_x\times 1}$ are their corresponding beamforming vectors, respectively.
	Therefore, the covariance matrix of the transmit signal can be defined as
	\begin{equation}
		\mathbf{R_x}=\mathbb{E}\left\lbrace \mathbf{x}(l)\mathbf{x}^{\mathrm{H}}(l)\right\rbrace
		\label{cov}
	\end{equation}
	and {maximum} transmit power at the BS can be calculated as
	\begin{align}
		&P_{\text{BS}}\geq\sum_{p=1}^{K}\mathbf{w}^{\mathrm{H}}_{c,p}\mathbf{w}_{c,p}+\sum_{q=1}^{M}\mathbf{w}_{s,q}^{\mathrm{H}}\mathbf{w}_{s,q} \nonumber\\
		&\hspace{0.5cm}\geq\Tr(\mathbf{R_x}).
		\label{cov}
	\end{align}
	
	
	In the proposed scheme, communication signals are generated as complex Gaussian random variables with zero mean and unit variance, i.e., $\mathbb{E}\left\lbrace \mathbf{c}(l)\mathbf{c}(l)^{\mathrm{H}}\right\rbrace =\mathbf{I}_K$, while sensing signals  are generated as frequency modulated continuous wave (FMCW) signals  \cite{de2021joint} as  follows
	\begin{equation}
		{s}_q(l)=\sum_{\tau=1}^{L_c}{A_\tau}\rect\bigg( \frac{l-\left( \tau-\frac{1}{2}T_c\right) }{T_c}\bigg) \cos\bigg( 2\pi f_c l + \pi \frac{B_{w}}{L}l^2\bigg) 
	\end{equation}
	where {$L_c$ is the number of chips, $L$ is the chirp length, $T_c=L/L_c$ is the chip duration}, the  $A_\tau$ is the amplitude of the signal at the $\tau$-th time block, $A_\tau\in\left\lbrace -1, 1 \right\rbrace $, $f_c$ is the carrier frequency and $B_w$ is the bandwidth. {Moreover, it is assumed that communication and sensing signals are mutually uncorrelated  i.e., } $\mathbb{E}\left\lbrace \mathbf{c}(l)\mathbf{s}(l)^{\mathrm{H}}\right\rbrace =\mathbf{0}_{K\times M}$.

	\subsection{Communication Model}
	In the proposed system, all communication channels  are  modeled by Rician fading channels, and the channels  between BS-STAR-RIS (BS-RIS), BS-User $k$ at Side-$R$ ($\text{BS-U}_{k, R}$) and STAR-RIS-User $k$ at Side $i$ ($\text{RIS-U}_{k, i}$) are respectively represented by $\mathbf{G}\in\mathbb{C}^{N\times T_x}$, $\mathbf{g}_{k, R}\in\mathbb{C}^{1\times T_x} $ and $\mathbf{h}_{k, i}\in\mathbb{C}^{1\times  N}$,   where $i\in\mathcal{S}=\left\lbrace T, R\right\rbrace $  and $k \in\left\lbrace 1, \cdots, K_i \right\rbrace $. {Here, to simplify the presentation, a unified notation $\mathbf{D}_c$ is introduced, which can  represent 
	 $\mathbf{D}_c\in\left\lbrace \mathbf{G}, \mathbf{g}_{k,R}, \mathbf{h}_{k, i} \right\rbrace $, as follows:} 
	\begin{align}
		&	\mathbf{D}_c=\sqrt{{\alpha}_{c}}\left( \sqrt{\frac{\kappa}{1+\kappa}}\mathbf{D}_c^{\text{LOS}}+\sqrt{\frac{1}{1+\kappa}}\mathbf{D}_c^{\text{NLOS}}\right) 
		\label{channel}
	\end{align}
	where $\mathbf{D}_c^{\text{LOS}}$  represents the LOS component, $\mathbf{D}_c^{\text{NLOS}}$ stands for the  NLOS component, $\kappa$ is the Rician factor and   $\alpha_c={\alpha_0} d_{c}^{-\rho} \in\big\{ \alpha_{\text{BS-RIS}}, \alpha_{\text{BS-U}_{k, R}}, \alpha_{\text{RIS-U}_{k, i}} \big\} $  denotes corresponding path attenuation, for $\alpha_0$ being the reference path loss at 1 meter (m), $\rho$  being  path-loss exponent and  $d_c\in\left\lbrace {d_{\text{BS-RIS}}, d_{\text{BS-U}_{k, R}}, d_{\text{RIS-U}_{k, i}}}\right\rbrace $ being the corresponding distance. Here, each element of $\mathbf{D}_c^{\text{NLOS}}$ component  is assumed to be generated by independent and identically distributed (i.i.d.) complex Gaussian random variables  following $\mathcal{CN}(0,1)$ distribution, while $\mathbf{D}_c^{\text{LOS}}$ component is deterministic channel generated by steering vectors. {Therefore, the overall  $\mathbf{D}_c$ channel follows $\mathcal{CN}(\sqrt{{\alpha}_{c}}\sqrt{\frac{\kappa}{1+\kappa}}\mathbf{D}_c^{\text{LOS}}, {\alpha}_{c}\frac{1}{1+\kappa}\mathbf{I} )$}.
	
	Using steering vectors, the LOS component $\mathbf{G}^{\text{LOS}}\in\mathbb{C}^{N\times T_x}$ of the channel between BS-STAR-RIS  is generated as: 
	\begin{equation}
		\vspace{0.00cm}	\mathbf{G}^{\text{LOS}}=\mathbf{b}^{\mathrm{H}}(\varphi^{\text{RIS}}_{c,h} ,\varphi^{\text{RIS}}_{c,v})\mathbf{a}(\theta^{\text{RIS}}_{c,h},\theta^{\text{RIS}}_{c,v})
	\end{equation}
	where $\mathbf{a}(\theta^{\text{RIS}}_{c,h},\theta^{\text{RIS}}_{c,v})\in\mathbb{C}^{1\times T_x}$  and $\mathbf{b}(\varphi^{\text{RIS}}_{c,h} ,\varphi^{\text{RIS}}_{c,v})\in\mathbb{C}^{1\times N}$ are the ULA and UPA array factors of BS and hybrid STAR-RIS, respectively.  Here, $N=N_x\times N_y$ with $N_x$ and $N_y$ representing  the number of hybrid STAR-RIS  elements  located horizontally and vertically on $x$-$y$ plane, respectively. $\theta^{\text{RIS}}_{c,h},\theta^{\text{RIS}}_{c,v}$ are horizontal and vertical AoDs  from BS towards RIS,  while $\varphi^{\text{RIS}}_{c,h} ,\varphi^{\text{RIS}}_{c,v}$ are the horizontal, vertical angles of arrival (AoA)  at RIS, respectively.  Here, $\mathbf{a}(\theta^{\text{RIS}}_{c,h},\theta^{\text{RIS}}_{c,v})$ and $\mathbf{b}(\varphi^{\text{RIS}}_{c,h} ,\varphi^{\text{RIS}}_{c,v})$ array factors are given as 
	\begin{align}
		&\mathbf{a}(\theta^{\text{RIS}}_{c,h},\theta^{\text{RIS}}_{c,v})=[1,\cdots,e^{j\eta_{\text{BS}}(T_x-1)\sin(\theta^{\text{RIS}}_{c,h})\cos(\theta^{\text{RIS}}_{c,v})}]\label{bs}\\
		&\mathbf{b}(\varphi^{\text{RIS}}_{c,h} ,\varphi^{\text{RIS}}_{c,v})=\mathbf{b}^{x}(\varphi^{\text{RIS}}_{c,h} ,\varphi^{\text{RIS}}_{c,v})\otimes\mathbf{b}^{y}(\varphi^{\text{RIS}}_{c,h} ,\varphi^{\text{RIS}}_{c,v})\label{ris}
	\end{align}
	where $\eta_{\text{BS}}=\frac{2\pi }{\lambda}d_{\text{BS}}$ for $\lambda$ wavelength and $d_{\text{BS}}$ distance between two adjacent transmit/receive antennas at BS, respectively.  Here, $\mathbf{b}^{x}(\varphi^{\text{RIS}}_{c,h},\varphi^{\text{RIS}}_{c,v})\in\mathbb{C}^{1\times N_x}$ and $\mathbf{b}^{y}(\varphi^{\text{RIS}}_{c,h} ,\varphi^{\text{RIS}}_{c,v})\in\mathbb{C}^{1\times N_y}$ are respectively steering vectors of STAR-RIS in $x$ and $y$ dimensions given as
	\begin{align}
		&\mathbf{b}^{x}(\varphi^{\text{RIS}}_{c,h} ,\varphi^{\text{RIS}}_{c,v})=[1,\cdots,e^{j\eta_{\text{RIS}}(N_x-1)\sin(\varphi^{\text{RIS}}_{c,h})\cos(\varphi^{\text{RIS}}_{c,v})}]\\
		&\mathbf{b}^{y}(\varphi^{\text{RIS}}_{c,h} ,\varphi^{\text{RIS}}_{c,v})=[1, \cdots, e^{j\eta_{\text{RIS}}(N_y-1)\sin(\varphi^{\text{RIS}}_{c,h})\sin(\varphi^{\text{RIS}}_{c,v})}]
	\end{align}
	where $\eta_{\text{RIS}}=\frac{2\pi }{\lambda}d_{\text{RIS}}$ for  $d_{\text{RIS}}$ being horizontal and vertical distance between two adjacent elements at STAR-RIS.  Therefore, (\ref{ris}) can be simply rewritten as 
	\begin{align}
		\mathbf{b}(\varphi^{\text{RIS}}_{c,h} ,\varphi^{\text{RIS}}_{c,v})=e^{j\eta_{\text{RIS}}\mathbf{k}(\varphi_{c,h}^{\text{RIS}},\varphi_{c,v}^{\text{RIS}})}
		\label{bb}
	\end{align}
	where $\mathbf{k}(\varphi_{c,h}^{\text{RIS}},\varphi_{c,v}^{\text{RIS}})\in\mathbb{C}^{1\times N}=\mathbf{k}_x\sin(\varphi^{\text{RIS}}_{c,h})\cos(\varphi^{\text{RIS}}_{c,v})+\mathbf{k}_y\sin(\varphi^{\text{RIS}}_{c,h})\sin(\varphi^{\text{RIS}}_{c,v})$. Here, $\mathbf{k}_x\in\mathbb{C}^{1\times N}=\mathbf{e}_x\otimes\mathbf{1}_{N_y}$ and \mbox{$\mathbf{k}_y\in\mathbb{C}^{1\times N}=\mathbf{1}_{N_y}\otimes\mathbf{e}_x$}  are with scalar components for $\mathbf{e}_x\in\mathbb{C}^{1\times N_x}=[0,1\cdots,N_x-1]$ and \mbox{$\mathbf{e}_y\in\mathbb{C}^{1\times N_y}=[0,1\cdots,N_y-1]$}.
	
	Similarly, the LOS channel components of the channel between BS-U$_{k,R}$ and RIS-U$_{k_i}$ can be generated as  $\mathbf{g}_{k, R}^{\text{LOS}}=\mathbf{a}(\theta_{c,h}^{k,R}, \theta^{k,R}_{c,v})\in\mathbb{C}^{1\times T_x}$ and \mbox{$\mathbf{h}_{k,i}^{\text{LOS}}=\mathbf{b}(\varphi^{k,i}_{c,h} ,\varphi^{k,i}_{c,v})\in\mathbb{C}^{1\times N}$}, where $\big\{ \theta_{c,h}^{k,R},\theta^{k,R}_{c,v}\big\} $  and $\big\{\varphi^{k,i}_{c,h} ,\varphi^{k,i}_{c,v}\big\}$ are horizontal and vertical  AoD pairs from the BS towards U$_{k,R}$ and  from STAR-RIS towards U$_{k,i}$ at Side  $i\in\mathcal{S}=\left\lbrace T, R \right\rbrace$, respectively. Then, following ULA and UPA in (\ref{bs}) and (\ref{bb}) $\mathbf{a}(\theta_{c,h}^{k,R}, \theta^{k,R}_{c,v})$ and $\mathbf{b}(\varphi^{k,i}_{c,h} ,\varphi^{k,i}_{c,v})$ can be given as
	\begin{align}
		&\mathbf{a}(\theta_{c,h}^{k,R},\theta^{k,R}_{c,})=[ 1,\cdots,e^{j\eta_{\text{BS}}(T_x-1)\sin(\theta_{c,h}^{k,R})\cos(\theta_{c,v}^{k,{R}})}]  \label{direct-k}\\
		&\mathbf{b}(\varphi^{k,i}_{c,h} ,\varphi^{k,i}_{c,v})=e^{j\eta_{\text{RIS}}\mathbf{k}(\varphi_{c,h}^{k,i},\varphi_{c,v}^{k,i})}
	\end{align}
	where 
	$\mathbf{k}(\varphi_{c,h}^{k,i},\varphi_{c,v}^{k,i})\in\mathbb{C}^{1\times N}=\mathbf{k}_x\sin(\varphi_{c,h}^{k,i})\cos(\varphi_{c,v}^{k,i})+\mathbf{k}_y\sin(\varphi_{c,h}^{k,i})\sin(\varphi_{c,v}^{k,i})$.
	Then, following (\ref{beam1}) and (\ref{channel}),  the received signal at U$_{k,T}$ that perceives  the transmit signals from the BS over the the active transmissive elements of the hybrid STAR-RIS becomes 
{\begin{align}
		&{y}^{k,T}_c(l)=\mathbf{h}_{k,T}\mathbf{\Phi}_t\mathbf{G}\mathbf{w}_{c,k}c_k(l)+\underbrace{\sum_{q=1}^{M}\mathbf{h}_{k,T}\mathbf{\Phi}_t\mathbf{G}\mathbf{w}_{s,{q}}s_q(l)}_{\text{\scriptsize  
				{sensing interference}}}\nonumber\\
		&\resizebox{1\columnwidth}{!}{$+\underbrace{\sum_{\tilde{k} \neq k}^{K_T} \mathbf{h}_{k,T} \mathbf{\Phi}_t \mathbf{G} \mathbf{w}_{c,\tilde{k}} c_{\tilde{k}}(l) + \sum_{\tilde{k} = K_T+1}^{K} \mathbf{h}_{k,T} \mathbf{\Phi}_t \mathbf{G} \mathbf{w}_{c,\tilde{k}} c_{\tilde{k}}(l)
		}_{\text{user interference}} +\underbrace{\mystrut{1ex}\mathbf{h}_{k,T}\mathbf{\Phi}_t\mathbf{v}_1}_{\text {thermal noise}}+\underbrace{\mystrut{1ex}{n}^{k,T}_c(l)}_{\text{ static noise}}$}
			\label{received-kt}
\end{align}}where ${n}^{k,T}_c(l)\in\mathcal{CN}(0,\sigma_{n_c}^2)$ is the static additive white Gaussian  noise  (AWGN) {sample}  and  $\mathbf{v}_1\in\mathbb{C}^{N\times 1} $ is the { non-negligible internal thermal noise introduced by  reflection-type amplifiers of active STAR-RIS elements \cite{zhi2022active}} whose each element is i.i.d. and follows $\mathcal{CN}({0},\sigma_{v_1}^2)$  for $k\in\left\lbrace 1,\cdots,K_T \right\rbrace $ and $\tilde{k}\in\left\lbrace 1,\cdots,K \right\rbrace $. 
	
	Likewise, for the communication users at Side-$R$ that receive  signals from BS through the direct path and reflected path over passive reflective elements of the hybrid STAR-RIS, the received signal at U$_{k,R}$  can be given as
{\begin{align}
		&{y}^{k,R}_c(l)=\nonumber\\
	&\resizebox{0.95\columnwidth}{!}{$\bar{\mathbf{h}}_{k,R}\mathbf{w}_{c,k}c_k(l)+\sum\limits_{\tilde{k}\neq k}^{K_R}\bar{\mathbf{h}}_{k,R}\mathbf{w}_{c,\tilde{k}}c_{\tilde{k}}(l)+\sum\limits_{\tilde{k}=K_R+1}^{K}\bar{\mathbf{h}}_{k,R}\mathbf{w}_{c,\tilde{k}}c_{\tilde{k}}(l)$}\nonumber\\
	&\hspace{0.0cm}+{\sum_{q=1}^{M}\bar{\mathbf{h}}_{k,R}\mathbf{w}_{s,{q}}s_q(l)}+ {n}^{k,R}_c(l)
		\label{received-kr}	\end{align}}where  $\bar{\mathbf{h}}_{k,R}=\mathbf{h}_{k,R}\mathbf{\Phi}_r\mathbf{G}+\mathbf{g}_{k,R}$ and ${n}^{k,R}_c(l)\sim\mathcal{CN}(0,\sigma_{n_c}^2)$ is the overall noise figure between the BS and U$_{k,R}$ for $k\in\left\lbrace 1,\cdots,K_R \right\rbrace $ and $\tilde{k}\in\left\lbrace 1,\cdots,K \right\rbrace $. 

	
	
	\subsection{Sensing Model}
	In the proposed system, the paths from the BS and hybrid STAR-RIS towards the targets are  deterministic channels and constructed  using steering vectors.  
	The channel between 
	the STAR-RIS and the target $m$ at Side-$i$ (O$_{m,i}$) can be given in terms of  the AoD pair of $\big\{\varphi^{{m,i}}_{h} ,\varphi^{m,i}_{v}\big\}$ as follows 
	\begin{align}
		&\mathbf{b}(\varphi^{{m,i}}_{s,h} , \varphi^{{m,i}}_{s,v})=e^{j\eta_{\text{RIS}}\mathbf{k}(\varphi^{{m,i}}_{s,h},\varphi^{{m,i}}_{s,v})}\label{b1}
	\end{align}
	where $\mathbf{b}(\varphi^{{m,i}}_{s,h} ,\varphi^{{m,i}}_{s,v})\in\mathbb{C}^{1\times  N}$, $\mathbf{k}(\varphi^{{m,i}}_{s,h},\varphi^{{m,i}}_{s,v})\in\mathbb{C}^{1\times N}=\mathbf{k}_x\sin(\varphi^{{m,i}}_{s,h})\cos(\varphi^{{m,i}}_{s,v})+\mathbf{k}_y\sin(\varphi^{{m,i}}_{s,h})\sin(\varphi^{{m,i}}_{s,v})$. 
	Thus, the received echo signal at the BS from the $m$th sensing target on Side-$T$  (O$_{m,T}$) follows the path of BS$\rightarrow$STAR-RIS$\rightarrow$O$_{m,T}\hspace{-0.1cm}\rightarrow$STAR-RIS$\rightarrow$BS. In order to mitigate severe path attenuation due to this multi-hop transmission, the considered dual-sided hybrid STAR-RIS utilizes active elements in transmission mode in both directions.   Then, the received echo signal collected at the BS due to the sensing target O$_{m,T}$ becomes
	\begin{align}
		&\mathbf{y}_s^{m,T}(l) =\beta_{\text{RIS}}^{m,T} \mathbf{G}^{\mathrm{H}}\mathbf{\Phi}^{\mathrm{H}}_t\mathbf{B}_m\mathbf{\Phi}_t\mathbf{G}\mathbf{x}(l)+\tilde{\mathbf{n}}_s^{m,T}(l)
		\label{targetB}
	\end{align}where  $m\in\left\lbrace 1,\cdots,M_T\right\rbrace $, $\beta_{\text{RIS}}^{m,T}$ is the round-trip path attenuation,  \mbox{$\mathbf{B}_m=\mathbf{b}^{\mathrm{H}}(\varphi_{s,h}^{m,T},\varphi_{s,v}^{m,T})\mathbf{b}(\varphi_{s,h}^{m,T},\varphi_{s,v}^{m,T})$}, {$\tilde{\mathbf{n}}_s^{m,T}(l)={\mathbf{G}^{\mathrm{H}}\mathbf{\Phi}^{\mathrm{H}}_t\mathbf{B}_m\mathbf{\Phi}_t\mathbf{v}_1(l) +\mathbf{G}^{\mathrm{H}}\mathbf{\Phi}^{\mathrm{H}}_t\mathbf{v}_2}(l)+\mathbf{n}_s^{m,T}(l)$} is the overall noise {samples} vector including thermal noise of active transmissive elements and static noise in the channel for $\mathbf{n}_s^{m,T}(l)$ being static noise and following $\sim\mathcal{CN}(\mathbf{0},\sigma_{n_s}^2\mathbf{I})$ distribution, while $\mathbf{v}_1(l)$ and $\mathbf{v}_2(l)$ are the thermal noise vectors with the $\mathcal{CN}(\mathbf{0},\sigma_{v_1}^2\mathbf{I})$ and $\mathcal{CN}(\mathbf{0},\sigma_{v_2}^2\mathbf{I})$ distributions, respectively. Here, for the sake of simplicity, the thermal and static noise variances are assumed to be  $\sigma_{n_s}^2=\sigma_{v,1}^2=\sigma_{v,2}^2$. 
	{It is important to note that  for $P_A=\sqrt{\Tr(\mathbf{\Phi}_t^{\mathrm{H}}\mathbf{\Phi}_t)}=\sqrt{\sum_{n=1}^{N}\left| \xi_{t,n} \right|^2}$ being the overall amplification factor of the active transmissive elements,   a low $P_A$ amplification factor implies that the internal thermal noise of the active transmissive STAR-RIS elements remains negligible. This is because the path attenuation terms in the first and second  expressions of $\tilde{\mathbf{n}}_s^{m,T}(l)$ make them insignificant, resulting in  $\tilde{\mathbf{n}}_s^{m,T}(l)\approx\mathbf{n}_s^{m,T}(l)$.  In this case,   the distribution of $\tilde{\mathbf{n}}_s^{m,T}(l)$ approximates to $\sim\mathcal{CN}(\mathbf{0},\sigma_{\tilde{n}_s}^2\mathbf{I})$ where $\sigma_{\tilde{n}_s}^2= \sigma_{n_s}^2$. On the other hand, for very high $P_A$ values,  the internal thermal noise effect becomes  significant. In that case, $\tilde{\mathbf{n}}_s^{m,T}(l)\approx\mathbf{G}^{\mathrm{H}}\mathbf{\Phi}^{\mathrm{şH}}_t\mathbf{B}_m\mathbf{\Phi}_t\mathbf{v}_1(l) $. Thus, the overall noise variance becomes
		$
		\sigma_{\tilde{n}_s}^2 = \sigma_{{n}_s}^2 \left| (\mathbf{\Psi}({n,n} ) \right|^2 \alpha_{\text{BS-RIS}} \left( \frac{\kappa}{1+\kappa} \sum\limits_{\upsilon} |\mathbf{G}^{\text{LOS}}({n},{\upsilon})|^2 + \frac{1}{1+\kappa} \right)$, where  $\mathbf{\Psi}=\mathbf{\Phi}^{\mathrm{H}}_t\mathbf{B}_m\mathbf{\Phi}_t$ and $\upsilon\in\left\lbrace 1,2,\cdots, T_x\right\rbrace $.{\addtocounter{footnote}{-1}\let\thefootnote\svthefootnote\footnote[2]{For \( Z = XY \), where \( X \) and \( Y \) are independent random variables, the distribution of \( Z \) is characterized by  \( \mu_Z = \mu_X \mu_Y \) and \( \sigma_Z^2 = \sigma_X^2 \mu_Y^2 + \sigma_Y^2 \mu_X^2 + \mu_X^2 \mu_Y^2 \) {\cite{proakis2007fundamentals}}.}}}

	
	Then,  using (\ref{targetB}), the overall received echo signal over the $L$ time block is given  as  
	\begin{align}
		&\hspace{-0.3cm}	\mathbf{Y}_s^{m,T} = \beta_{\text{RIS}}^{m,T} \mathbf{G}^{\mathrm{H}}\mathbf{\Phi}^{\mathrm{H}}_t\mathbf{B}_m\mathbf{\Phi}_t\mathbf{G}\mathbf{X}+\tilde{\mathbf{N}}_s^{m,T}
		\label{receivedt}
	\end{align} 
	where  $\mathbf{Y}_s^{m,T}=[\mathbf{y}_s^{m,T}(1), \cdots, \mathbf{y}_s^{m,T}(L)]$, $\mathbf{X}=[\mathbf{x}(1), \cdots, \mathbf{x}(L)]$ and $\tilde{\mathbf{N}}_s^{m,T}=[\tilde{\mathbf{n}}_s^{m,T}(1), \cdots, \tilde{\mathbf{n}}_s^{m,T}(L)]$.
	\begin{figure*}[t!]
		\begin{align}\tag{31}
			\gamma_{m,T}\leq\frac{(\beta_{\text{RIS}}^{m,T})^2\Tr( \bar{\mathbf{G}}_{}^{}\mathbf{\Phi}_t^{}\mathbf{B}_m^{}\mathbf{\Phi}_t^{\mathrm{H}})^2\Tr(\mathbf{R_x}) }{\sum\limits_{j\neq m}^{M_R}  \Tr (\mathbf{f}_{j,R}^{\mathrm{H}} \mathbf{f}_{j,R}^{})^2\Tr(\mathbf{R_x}) + \sum\limits_{\tilde{m}\neq m}^{M_T}(\beta_{\text{RIS}}^{\tilde{m},T})^2\Tr (\bar{\mathbf{G}}\mathbf{\Phi}_t^{}\mathbf{B}_{\tilde{m},T}\mathbf{\Phi}_t^{\mathrm{H}})^2\Tr(\mathbf{R_x}) +{\sigma_{\tilde{n}_s}^2}}
			\label{gamma}
		\end{align}	
		\noindent\hrulefill
	\end{figure*}
	\begin{figure*}[t]
		\begin{align}\tag{32}
	&\gamma_{m,T}\leq\frac{(\beta_{\text{RIS}}^{m,T})^2\Tr (\bar{\mathbf{B}}_{m,T}^{}\mathbf{\Phi}_t^{\mathrm{H}}\mathbf{A}_{}^{}\mathbf{\Phi}_t^{})\Tr(\mathbf{\Phi}_t^{}\mathbf{\Phi}_t^{\mathrm{H}})\Tr(\mathbf{R_x}) }{\sum\limits_{j\neq m}^{M_R}  \Tr(\mathbf{f}_{j,R}^{\mathrm{H}} \mathbf{f}_{j,R}^{})^2 \Tr(\mathbf{R_x})+ \sum\limits_{\tilde{m}\neq m}^{M_T}(\beta_{\text{RIS}}^{\tilde{m},T})^2\Tr(\bar{\mathbf{B}}_{\tilde{m},T}^{\mathrm{H}}\mathbf{\Phi}_t^{\mathrm{H}}{\mathbf{A}_{}^{}}\mathbf{\Phi}_t^{})\Tr(\mathbf{\Phi}_t^{}\mathbf{\Phi}_t^{\mathrm{H}})\Tr(\mathbf{R_x}) +{{\sigma_{\tilde{n}_s}^2}}}
			\label{gamma1}
		\end{align}	
		\noindent\hrulefill
	\end{figure*}
	In the proposed scheme,   the active transmissive STAR-RIS elements are first employed  to transmit BS signals to  Side-$T$ and then reflect the echo signals of  the {distant} targets  back to the BS. Therefore, using (\ref{receivedt}), the overall power consumption of the active transmissive elements can be calculated as
	\begin{equation}
		\begin{aligned}
			&\hspace{-0.0cm}P_{\text{RIS}}\geq\\
			&\hspace{0cm}\resizebox{1\columnwidth}{!}{$\Big\|\mathbf{\Phi}_t\mathbf{Gx}(l) \Big\|^2+\left\|  \mathbf{\Phi}^{\mathrm{H}}_t\mathbf{B}_m\mathbf{\Phi}_t\mathbf{Gx}(l) \right\|^2+ \left\| \mathbf{\Phi}^{\mathrm{H}}_t\mathbf{B}_m\mathbf{\Phi}_t\mathbf{v}_1(l)\right\|^2+\left\| {\mathbf{\Phi}^{\mathrm{H}}_t\mathbf{v}_2}(l) \right\|^2$}.\label{power}
		\end{aligned}
	\end{equation}
	After some  manipulations, (\ref{power}) can be rewritten as
		\begin{align}
			&\hspace{-0.2cm}P_{\text{RIS}}\geq\Tr (\mathbf{\Phi}_t\mathbf{GR_x} \mathbf{G}^{\mathrm{H}}\mathbf{\Phi}^{\mathrm{H}}_t)+\Tr (\mathbf{\Phi}^{\mathrm{H}}_t\mathbf{B}_m\mathbf{\Phi}_t\mathbf{GR_x} \mathbf{G}^{\mathrm{H}}\mathbf{\Phi}^{\mathrm{H}}_t\mathbf{B}_m^{\mathrm{H}}\mathbf{\Phi}_t)\nonumber \\ &\hspace{1.5cm}+\sigma_{v_1}^2\Tr(\mathbf{\Phi}^{\mathrm{H}}_t\mathbf{B}_m\mathbf{\Phi}_t\mathbf{\Phi}^{\mathrm{H}}_t\mathbf{B}_m^{\mathrm{H}}\mathbf{\Phi}_t)+ \sigma_{v_2}^2\Tr(\mathbf{\Phi}^{\mathrm{H}}_t\mathbf{\Phi}_t). 
			\label{pris}
		\end{align}
		
		Furthermore, for the mono-static MIMO radar systems whose transmit and receive elements are co-located, the round-trip path loss is calculated as \cite{shao2022target, yu2023active}
		\begin{align}
			\beta= \sqrt{\frac{\lambda^2\Lambda}{(4\pi)^3d^4}}
			\label{rcs}
		\end{align}
		where $\Lambda$ is the radar cross section (RCS) and $d$ is the distance between MIMO radar and the sensing target. Therefore, in the proposed scenario, since the hybrid STAR-RIS can be considered as a mono-static MIMO radar \cite{yu2023active},   the path attenuation coefficient of $\beta_{\text{RIS}}^{m,T}$ in (\ref{receivedt}) is calculated for the corresponding  distance $d=d_{\text{RIS-O}_{m,T}}$. 
		
		Unlike the {distant} sensing targets,  since the {local} sensing targets on Side-$R$ have both   direct and indirect connections with the BS, similar to   (\ref{direct-k}),  the channel between BS and sensing target $m$ on Side-$R$ (O$_{m, R}$), $\mathbf{a}(\theta^{{m,R}}_{s,h},\theta^{{m,R}}_{s,v})\in\mathbb{C}^{1\times T_x}$,  can be expressed as 
		\begin{align}
			&\mathbf{a}(\theta^{{m,R}}_{s,h} ,_{}^{}\theta^{{m,R}}_{s,v})=[1,\cdots,e^{j\eta_{\text{BS}}(T_x-1)\sin(\theta^{m,R}_{s,h})\cos(\theta^{m,R}_{s,v})}]
			\label{a1}
		\end{align}
		where $m\in\left\lbrace 1,\cdots,M_R\right\rbrace $, $\theta^{m,R}_{s,h}$ and $\theta^{m,R}_{s,v}$ are horizontal and vertical AoDs from BS towards target O$_{m,R}$, respectively. Therefore, following (\ref{a1}) and (\ref{b1}) for $i=R$, the received echo signal collected at the BS from the target O$_{m,R}$ over the paths of BS$\rightarrow$O$_{m,R}\hspace{-0.12cm}\rightarrow$BS  and  BS$\hspace{-0.00cm}\rightarrow$STAR-RIS$\rightarrow$O$_{m,R}\hspace{-0.13cm}\rightarrow$STAR-RIS$\rightarrow$BS   becomes \cite{meng2023ris}
		{\small
			\begin{align}
				&	\mathbf{y}_s^{m, R}(l)={\beta^{m,R}_{\text{BS}}}\big(\mathbf{a}(\theta^{{m,R}}_{s,h},\theta^{{m,R}}_{s,v})+\tilde{\mathbf{b}}(\varphi^{{m,R}}_{s,h} ,\varphi^{{m,R}}_{s,v})\mathbf{\Phi}_r\mathbf{G}\big)^{\text{H}}\nonumber\\&\hspace{1cm}\times\big(\mathbf{a}(\theta^{{m,R}}_{s,h},\theta^{{m,R}}_{s,v})+\tilde{\mathbf{b}}(\varphi^{{m,R}}_{s,h} ,\varphi^{{m,R}}_{s,v})\mathbf{\Phi}_r\mathbf{G}\big)\mathbf{x}(l)+\mathbf{n}_s^{m,R}(l)
				\label{targetA}
		\end{align}}where $\mathbf{n}_s^{m,R}$ is the AWGN matrix whose each entry is i.i.d and with $\sim\mathcal{CN}(0,\sigma_{m,R}^2)$ distribution and \sloppy \mbox{$\tilde{\mathbf{b}}(\varphi^{{m,R}}_{s,h},\varphi^{{m,R}}_{s,v})\hspace{0.1cm}\in\hspace{0.1cm}\mathbb{C}^{1\times N} =\sqrt{{\beta^{m,R}_{\text{RIS}}}/{\beta^{m,R}_{\text{BS}}}}\hspace{0.1cm}\mathbf{b}(\varphi^{{m,R}}_{s,h} ,\varphi^{{m,R}}_{s,v})$ for   } $\beta^{m,R}_{\text{BS}}$ and $\beta^{m,R}_{\text{RIS}}$  being the round-trip path attenuation from the BS and hybrid STAR-RIS to the target O$_{m,R}$ calculated as in (\ref{rcs}) for corresponding $d=d_{\text{BS-O}_{m,R}}$ and $d=d_{\text{RIS-O}_{m,R}}$, respectively.  Then, over the $L$   time block,  (\ref{targetA}) can be rewritten as
		{\small
			\begin{align}
				&	\mathbf{Y}_s^{m, R}={\beta^{m,R}_{\text{BS}}}\big(\mathbf{a}(\theta^{{m,R}}_{s,h},\theta^{{m,R}}_{s,v})+\tilde{\mathbf{b}}(\varphi^{{m,R}}_{s,h} ,\varphi^{{m,R}}_{s,v})\mathbf{\Phi}_r\mathbf{G}\big)^{\text{H}}\nonumber\\&\hspace{1cm}\times\big(\mathbf{a}(\theta^{{m,R}}_{s,h},\theta^{{m,R}}_{s,v})+\tilde{\mathbf{b}}(\varphi^{{m,R}}_{s,h} ,\varphi^{{m,R}}_{s,v})\mathbf{\Phi}_r\mathbf{G}\big)\mathbf{X}+\mathbf{N}_s^{m,R}
				\label{rr}
		\end{align}}where   $\mathbf{Y}_s^{m,R}=[\mathbf{y}_s^{m,R}(1), \cdots, \mathbf{y}_s^{m,R}(L) ]$ and $\mathbf{N}_s^{m,R}=[\mathbf{N}_s^{m,R}(1), \cdots, \mathbf{N}_s^{m,R}(L) ]$. 

	\section{Performance Metrics and Problem Formulation}
	In this section, sensing and communication metrics of the   for the proposed system are derived. Subsequently,  an optimization problem that {maximizes  the minimum SINR } of the sensing targets, {subject to the communication SINR constraint, } is formulated  by  optimizing  transmissive/reflective  coefficients of the hybrid STAR-RIS and  beamforming matrix.    Moreover, the CRB for estimating 2D AoDs of the sensing targets is derived. {To effectively solve this optimization, the problem is reformulated into linear and quadratic components, thus  enabling the problem to be solved through a SDR-based approach.}


	\subsection{Target SINR Analysis}
	In this subsection, the performance of the target SINR  is determined. 
	
	Using (\ref{receivedt}), for the {distant} targets on the Side-$T$ that perceive transmit signals through the active transmissive STAR-RIS elements, the target SINR  can be given, considering  interference from the other sensing targets both on Side-$R$ and Side-$T$, as
	{\small
		\begin{equation}
			\begin{aligned}[b]
				&\gamma_{m,T} =\\
				&\frac{\left\| \beta_{\text{RIS}}^{m,T} \mathbf{G}^{\mathrm{H}}\mathbf{\Phi}^{\mathrm{H}}_t\mathbf{B}_m\mathbf{\Phi}_t\mathbf{G}\mathbf{X}\right\| ^2 }{\sum\limits_{j=1}^{M_R}\left\|  \mathbf{f}_{j,R}^{\mathrm{H}} \mathbf{f}_{j,R}^{}\mathbf{X}\right\|^2+ \sum\limits_{\tilde{m}\neq m}^{M_T} \left\|\beta_{\text{RIS}}^{\tilde{m},T} \mathbf{G}^{\mathrm{H}}\mathbf{\Phi}^{\mathrm{H}}_t\mathbf{B}_{\tilde{m},T}\mathbf{\Phi}_t\mathbf{G}\mathbf{X}\right\| ^2+{\sigma_{\tilde{n}_s}^2}}
				\label{Omb}
			\end{aligned}
	\end{equation}}where $\mathbf{f}_{j,R}^{}={\beta^{j,R}_{\text{BS}}}\big(\mathbf{a}(\theta^{{j,R}}_{s,h},\theta^{{j,R}}_{s,v})+\tilde{\mathbf{b}}(\varphi^{{j,R}}_{s,h} ,\varphi^{{j,R}}_{s,v})\mathbf{\Phi}_r\mathbf{G}\big)$.
	After some simplification{\addtocounter{footnote}{-1}\let\thefootnote\svthefootnote\footnote[2]{For $\mathbf{K}, \mathbf{L}$ and $\mathbf{M}$ being matrices with coherent dimensions, the trace of matrix product is $\Tr(\mathbf{KLM})=\Tr(\mathbf{MKL})=\Tr(\mathbf{LMK})$ \cite{zhang2017matrix}.}, (\ref{Omb}) can be rewritten as in \eqref{gamma} for $\bar{\mathbf{G}}=\mathbf{G}^{\mathrm{H}}\mathbf{G}$ at the beginning of the this page.


	Here, employing  Cauchy Schwartz inequality \cite{zhang2017matrix} for $\Tr (\mathbf{KL})^2\leq \Tr(\mathbf{K}^{\mathrm{H}}\mathbf{K})\Tr(\mathbf{L}^{\mathrm{H}}\mathbf{L})$, the target SINR of O$_{m,T}$ in (\ref{gamma}) is upper bounded as in (\ref{gamma1}). Then, using \cite[Theorem 1.10]{zhang2017matrix},  for $\mathbf{Q}_{m,T}\in\mathbb{C}^{N\times N}=\mathbf{B}_m\mathbf{B}_m^{\mathrm{H}}\odot\bar{\mathbf{G}}^{\mathrm{H}}\bar{\mathbf{G}}$, $\mathbf{Q}_{\tilde{m},T}\in\mathbb{C}^{N\times N}={\mathbf{B}}_{\tilde{m}}{\mathbf{B}}_{\tilde{m}}^{\mathrm{H}}\odot\bar{\mathbf{G}}^{\mathrm{H}}\bar{\mathbf{G}}$ and $\mathbf{Z}_t\in\mathbb{C}^{N\times N}=\mathbf{z}_t^\mathrm{H}\mathbf{z}_t^{}$ for $\mathbf{z}_t\in\mathbb{C}^{1\times N}=[\xi_{t,1} e^{j\phi_{t,1}},   \xi_{t,2}e^{j\phi_{t,2}},  \cdots,  \xi_{t,N} e^{j\phi_{t,N}}]$, the target SINR in (\ref{gamma1}) can be bounded as  in (\ref{sinrt}),  which appears at the top of the next page.
			\begin{figure*}[t]
			{{\begin{align}\tag{33}
						&\gamma_{m,T}\leq\frac{{P_{A}}^2(\beta_{\text{RIS}}^{m,T})^2\Tr(\mathbf{Q}_{m,T}\mathbf{Z}_t)\Tr(\mathbf{R_x})}{\sum\limits_{j\neq m}^{M_R}\Tr\left( (\mathbf{f}_{j,R}^{\mathrm{H}} \mathbf{f}_{j,R}^{})^2\right)\Tr(\mathbf{R_x}) + \sum\limits_{\tilde{m}\neq m}^{M_T}{P_A}^2(\beta_{\text{RIS}}^{\tilde{m},T})^2\Tr(\mathbf{Q}_{\tilde{m},T}\mathbf{Z}_t)\Tr(\mathbf{R_x}) +{\sigma_{\tilde{n}_s}^2}}
						\label{sinrt}
			\end{align}}}
			\noindent\hrulefill
		\end{figure*}
		\begin{figure*}[t!]
			\begin{align}\tag{35}
				&\gamma_{m,R} \leq\frac{(\beta_{\text{BS}}^{m,R})^2\Tr(\mathbf{R_x})\big(\Tr(\mathbf{A}_m^{}\mathbf{A}_m^{\mathrm{H}}) ^2 +N\Tr(\mathbf{Q}_{m,R}\mathbf{Z}_r)\big) }{\sum\limits_{\tilde{m}\neq m}^{M_R}(\beta_{\text{BS}}^{\tilde{m},R})^2\Tr(\mathbf{R_x})\big((\Tr(\mathbf{A}_{\tilde{m},R}^{}\mathbf{A}_{\tilde{m},R}^{\mathrm{H}}) ^2 +N\Tr(\mathbf{Q}_{\tilde{m},R}\mathbf{Z}_r)\big)+ \sum\limits_{j=1}^{M_T} {P_A}^2(\beta_{\text{RIS}}^{j,T})^2\Tr(\mathbf{Q}_{j,T}\mathbf{Z}_t)\Tr(\mathbf{R_x})+{\sigma_{{n}_s}^2}}
				\label{sinrr}
			\end{align}
			\noindent\hrulefill
		\end{figure*}
	
	Similarly, considering the received echo signal given in (\ref{rr}),  the SINR of the target O$_{m,R}$ can be expressed as 
\begin{align}\setcounter{equation}{33}
	&\gamma_{m,R}=\nonumber\\
	&\frac{\left\| \mathbf{f}_{k,R}^{\mathrm{H}}\mathbf{f}_{k,R}\mathbf{X}\right\|^2 }{\sum\limits_{\tilde{m}\neq m}^{M_R}\left\|  \mathbf{f}_{m,R}^{\mathrm{H}} \mathbf{f}_{m,R}^{}\mathbf{X}\right\|^2+ \sum\limits_{{j}= 1}^{M_T} \left\|\beta_{\text{RIS}}^{{j},T} \mathbf{G}^{\mathrm{H}}\mathbf{\Phi}^{\mathrm{H}}_t\mathbf{B}_{{j},T}\mathbf{\Phi}_t\mathbf{G}\mathbf{X}\right\| ^2+{\sigma_{{n}_s}^2}}.
\end{align}
Therefore, 	for $\Tr(\mathbf{\Phi}_r^{\mathrm{H}}\mathbf{\Phi}_r)=N$ and using $\Tr((\mathbf{K}+\mathbf{L})(\mathbf{K}+\mathbf{L})^{\mathrm{H}})\leq2\big(\Tr(\mathbf{K}\mathbf{K}^{\mathrm{H}})+\Tr(\mathbf{L}\mathbf{L}^{\mathrm{H}})\big)$ \cite{zhang2017matrix}, the target SINR of each O$_{m,R}$  can be upper bounded as in (\ref{sinrr}),  where $\mathbf{A}_m\in\mathbb{C}^{T_x\times T_x}=\mathbf{a}(\theta_{s,h}^{m,R},\theta_{s,v}^{m,R})\mathbf{a}^{\mathrm{H}}(\theta_{s,h}^{m,R},\theta_{s,v}^{m,R})$, $\mathbf{Q}_{{m},R} \in\mathbb{C}^{N\times N}= \tilde{\mathbf{B}}_{m}\tilde{\mathbf{B}}_{m}^{\mathrm{H}}\odot\bar{\mathbf{G}}^{\mathrm{H}}\bar{\mathbf{G}}$, $\tilde{\mathbf{B}}_{m}\in\mathbb{C}^{N\times N}=\tilde{\mathbf{b}}(\varphi^{{m,R}}_{s,h},\varphi^{{m,R}}_{s,v})^\mathrm{H}\tilde{\mathbf{b}}(\varphi^{{m,R}}_{s,h},\varphi^{{m,R}}_{s,v})$, $\mathbf{Z}_r\in\mathbb{C}^{N\times N}=\mathbf{z}_r^{\mathrm{H}}\mathbf{z}_r$, $\mathbf{z}_r\in\mathbb{C}^{1\times N}=[ e^{j\phi_{r,1}},  e^{j\phi_{r,2}},  \cdots,   e^{j\phi_{r,N}}]$ and $j\in\left\lbrace 1,\cdots,M_T\right\rbrace $.

	{\subsection{User SINR Analysis}
		
		In this subsection, the SINR of the communication users  is derived.
		
Using (\ref{received-kt}), the SINR of user U$_{k,T}$ can be simply given as
		{\small\begin{align}	\setcounter{equation}{35}
					&R_{k,T}=\nonumber\\
					&\frac{\left|  \mathbf{h}_{k,T}\mathbf{\Phi}_t\mathbf{G}\mathbf{w}_{c,k} \right|^2 }{\resizebox{1\columnwidth}{!}{$\sum\limits_{\tilde{k}\neq k}^{K_T}\left| \mathbf{h}_{k,T}\mathbf{\Phi}_t\mathbf{G}\mathbf{w}_{c,\tilde{k}}\right|^2+\sum\limits_{\tilde{k}=K_T+1}^{K}\left| \mathbf{h}_{k,T}\mathbf{\Phi}_t\mathbf{G}\mathbf{w}_{c,\tilde{k}}\right|^2+\sum\limits_{q=1}^{M}\left| \mathbf{h}_{k,T}\mathbf{\Phi}_t\mathbf{G}\mathbf{w}_{s,{q}}\right|^2+\sigma^2_{\tilde{n}_c}$}}\label{sinr_ut}
			\end{align}}where $\sigma^2_{\tilde{n}_c}$ is the variance of the  overall noise figure $\tilde{n}_c^{k,T}$ of  the user U$_{k,T}$ which is calculated as  $\sigma^2_{\tilde{n}_{c}}={P_A}^2\alpha_{{\text{RIS-U}_{k,T}}}\big({\frac{\kappa}{1+\kappa}}\big| \big| \mathbf{h}_{k,T}^{\text{LOS}}\big| \big| ^2+\frac{1}{1+\kappa}\big) $. 
		Here, using above  matrix conversions \cite{zhang2017matrix}, for $\mathbf{V}_{k,T}=\mathbf{h}_{k,T}^{\mathrm{H}}\mathbf{h}_{k,T}\odot(\mathbf{Gw}_{c,k})(\mathbf{Gw}_{c,k})^{\mathrm{H}}$,  $\mathbf{V}_{\tilde{k},T}=\mathbf{h}_{k,T}^{\mathrm{H}}\mathbf{h}_{k,T}\odot(\mathbf{Gw}_{c,\tilde{k}})(\mathbf{Gw}_{c,\tilde{k}})^{\mathrm{H}}$ and $\mathbf{V}_{q,T}=\mathbf{h}_{k,T}^{\mathrm{H}}\mathbf{h}_{k,T}\odot(\mathbf{Gw}_{c,q})(\mathbf{Gw}_{c,q})^{\mathrm{H}}$,  (\ref{sinr_ut}) can be rewritten in a simplified form as
		\begin{align}
		&R_{k,T}=\nonumber\\
		&\frac{\Tr(\mathbf{V}_{k,T}\mathbf{Z}_t)}{\sum\limits_{\tilde{k}\neq k}^{K_T}\Tr(\mathbf{V}_{\tilde{k},T}\mathbf{Z}_t)+\sum\limits_{\tilde{k}=K_T+1 }^{K}\Tr(\mathbf{V}_{\tilde{k},T}\mathbf{Z}_t)+\sum\limits_{{q}=1 }^{M}\Tr(\mathbf{V}_{{q},T}\mathbf{Z}_t)+\sigma^2_{\tilde{n}_c}}
		\end{align}
			
On the other hand, since  the communication user U$_{k,R}$ has a  direct link from the BS and the reflective passive elements of the hybrid STAR-RIS add a negligible contribution on the received signal determined in (\ref{received-kr}), the SINR of the user U$_{k,R}$ can be approximated as 
			{\small\begin{align}
					&R_{k,R}\approx\nonumber\\
					&\frac{\Tr\left(\tilde{\mathbf{V}}_{k,R}\right)}{{\sum\limits_{\tilde{k}\neq k}^{K_R}\Tr( \tilde{\mathbf{V}}_{\tilde{k},R})+\sum\limits_{\tilde{k}=K_R+1 }^{K}\Tr( \tilde{\mathbf{V}}_{\tilde{k},R})+\sum\limits_{q=1}^{M}{ \Tr( \tilde{\mathbf{V}}_{\tilde{q},R})}+\sigma^2_{c}}}
			\end{align}}where  $\tilde{\mathbf{V}}_{k,R}=(\mathbf{g}_{k,R}\mathbf{w}_{k,R})^{\mathrm{H}}(\mathbf{g}_{k,R}\mathbf{w}_{k,R})$, $\tilde{\mathbf{V}}_{\tilde{k},R}=(\mathbf{g}_{k,R}\mathbf{w}_{\tilde{k},R}),  ^{\mathrm{H}}(\mathbf{g}_{k,R}\mathbf{w}_{\tilde{k},R})$ and \mbox{$\tilde{\mathbf{V}}_{q,R}=(\mathbf{g}_{k,R}\mathbf{w}_{q,R}) ^{\mathrm{H}}(\mathbf{g}_{k,R}\mathbf{w}_{q,R})$}.}
		
			

		\subsection{Problem Formulation}
		To optimize the coefficients of the hybrid STAR-RIS  the following problem that maximizes the minimum target SINR under the communication SINR constraint is formulated. 

		\begin{subequations}
			\begin{align}
				&\text{(P1)}\hspace{0.2cm}{\max_{\mathbf{\Phi}_t, \mathbf{\Phi}_r}  \min_{m,\mathcal{S}}}\hspace{0.2cm}\gamma_{{m,i}}\\
				&\hspace{1.75cm}\text{s.t.} \hspace{0.3cm}\left| \mathbf{z}_r(n)\right| = 1,\hspace{0.2cm}\forall n\in\left\lbrace 1,\cdots,N\right\rbrace  \label{1a} \\
				&\hspace{2.45cm}\left| \mathbf{z}_t(n)\right| \geq 1,\hspace{0.2cm}\forall n\in\left\lbrace 1,\cdots,N\right\rbrace \label{1b}\\
				&\hspace{2.45cm} {R_{k,i}\geq R_{th}} \label{1c}\\\
				& \hspace{2.45cm} \Tr(\mathbf{R_x})\leq P_{\text{BS}}\label{1d}\\
				&\hspace{2.45cm}\Tr(\mathbf{Z}_t)\leq {P_A}^2.\label{1e}
			\end{align}
		\end{subequations}
	Here $R_{\text{th}}$ is the minimum SINR constraint for the communication users, {while $\mathbf{Z}_t = \mathbf{z}_t \mathbf{z}_t^\mathrm{H} \in \mathbb{C}^{N \times 1}$, subject to $\rank(\mathbf{Z}_t) = 1$ and $\mathbf{Z}_t \succeq 0$, and $\mathbf{Z}_r = \mathbf{z}_r \mathbf{z}_r^\mathrm{H} \in \mathbb{C}^{N \times 1}$, subject to $\rank(\mathbf{Z}_r) = 1$ and $\mathbf{Z}_r \succeq 0$, where $\mathbf{z}_t \in \mathbb{C}^{N \times 1}$ and $\mathbf{z}_r \in \mathbb{C}^{N \times 1}$, respectively.}
		Here,   it is obvious that problem (P1) is a non-convex QCQP problem, which can be transformed into a SDP problem using the semi-definite relaxation (SDR) approach via relaxing the constraints. Therefore, (P1) can be reformulated as 
		\begin{subequations}
		\begin{align}
			&\text{(P2)}\hspace{0.2cm}{\max_{\mathbf{\Phi}_t, \mathbf{\Phi}_r}  \min_{m,\mathcal{S}}}\hspace{0.2cm}\gamma_{{m,i}}\\
			&\hspace{1.75cm}\text{s.t.} \hspace{0.3cm} \text{ (\ref{1c}), (\ref{1d}), (\ref{1e}) }\\
				&\hspace{1.65cm} {\mathbf{Z}_r \succeq 0}\label{2a}\\
			& \hspace{1.65cm}{\mathbf{Z}_t \succeq 0}\label{2b}.
		\end{align}
	\end{subequations}
Next, (P2) can be reformulated as  a linear matrix inequality (LMI) as follows
{\begin{subequations}
	\begin{align}
		&\text{(P3)}\hspace{0.2cm}{\max_{\mathbf{\Phi}_t, \mathbf{\Phi}_r}  \min_{m,\mathcal{S}}}\hspace{0.2cm}
		\gamma_{m,i}\\
		&\hspace{0.8cm}\text{s.t.}\hspace{0.3cm}\resizebox{0.45\columnwidth}{!}{$\text{(\ref{1c}), (\ref{1d}), (\ref{1e}), (\ref{2a}), (\ref{2b})}$}\\
		&\hspace{0.5cm} \resizebox{0.9\columnwidth}{!}{$ \hspace{0.28cm}{P_{A}(\beta_{\text{RIS}}^{m,T})^2\Tr(\mathbf{Q}_{m,T}\mathbf{Z}_t)\Tr(\mathbf{R_x})}\geq \sum\limits_{j\neq m}^{M_R}\Tr\big( (\mathbf{f}_{j,R}^{\mathrm{H}} \mathbf{f}_{j,R}^{})^2\big)\Tr(\mathbf{R_x})$} \nonumber\\
		&\hspace{0.8cm}\resizebox{0.7\columnwidth}{!}{$+ \sum\limits_{\tilde{m}\neq m}^{M_T}{P_A}^2(\beta_{\text{RIS}}^{\tilde{m},T})^2\Tr(\mathbf{Q}_{\tilde{m},T}\mathbf{Z}_t)\Tr(\mathbf{R_x}) +{\sigma_{\tilde{n}_s}^2}$}\\
				&\hspace{0.9cm} \resizebox{0.8\columnwidth}{!}{$(\beta_{\text{BS}}^{m,R})^2\Tr(\mathbf{R_x})\big(\Tr(\mathbf{A}_m^{}\mathbf{A}_m^{\mathrm{H}}) ^2 +N\Tr(\mathbf{Q}_{m,R}\mathbf{Z}_r)\big)\geq$}\hspace{0.9cm}\nonumber\\
				&\hspace{0.9cm}\resizebox{0.8\columnwidth}{!}{$\sum\limits_{\tilde{m}\neq m}^{M_T}(\beta_{\text{RIS}}^{\tilde{m},T})^2\Tr(\bar{\mathbf{B}}_{\tilde{m},T}^{\mathrm{H}}\mathbf{\Phi}_t^{\mathrm{H}}{\mathbf{A}_{}^{}}\mathbf{\Phi}_t^{})\Tr(\mathbf{\Phi}_t^{}\mathbf{\Phi}_t^{\mathrm{H}})\Tr(\mathbf{R_x})$}\nonumber\\ 
					&\hspace{0.9cm}\resizebox{0.55\columnwidth}{!}{$+\sum\limits_{j\neq m}^{M_R}  \Tr(\mathbf{f}_{j,R}^{\mathrm{H}} \mathbf{f}_{j,R}^{})^2 \Tr(\mathbf{R_x})+{{\sigma_{\tilde{n}_s}^2}}$}.
	\end{align}	
\label{p3}
\end{subequations}}

	Here, the optimization problem (P3) can be efficiently solved with CVX solvers \cite{cvx}. {The complexity of solving the SDR problem (P3), which leverages the interior point method for a solution accuracy of $\epsilon>0$,  can be calculated as $\mathcal{O}((K^{4.5}M^{4.5}+N^{4.5})\log(1/\epsilon))$ \cite{luo2010semidefinite}}. {However, the SDR solution does not always guarantee the rank-one solution for the recovered $\hat{\mathbf{Z}}_t$ and $\hat{\mathbf{Z}}_r$ matrices. To address this, the eigenvalue decomposition \cite{wang2023stars} or Gaussian randomization \cite{luo2010semidefinite} can be employed to obtain a feasible rank-one solution to the problem (\ref{p3}). At this point, the eigenvalue decomposition  \cite{wang2023stars} is applied to recover the diagonal elements of $\hat{\mathbf{Z}}_t$ and $\hat{\mathbf{Z}}_r$ as  $\hat{\mathbf{z}}_t\in\mathbb{C}^{N\times 1}$ and $\hat{\mathbf{z}}_r\in\mathbb{C}^{N\times 1}$, respectively, where $\hat{\mathbf{Z}}_t=\hat{\mathbf{z}}_t\hat{\mathbf{z}}_t^\mathrm{H}$ and $\hat{\mathbf{Z}}_r=\hat{\mathbf{z}}_r\hat{\mathbf{z}}_r^\mathrm{H}$. }
	  
	Then the beamforming matrix is determined using optimal transmit beamforming via defining the overall beamforming matrix in (\ref{beam}) as $\mathbf{W}\in\mathbb{C}^{T_x\times(K+M)}=
	[\mathbf{w}_{c,1}, \cdots,\mathbf{w}_{c, K}, \mathbf{w}_{s,1}, \cdots,\mathbf{w}_{s, M}]$.  Here,  $\mathbf{w}_{c,p}$ and $\mathbf{w}_{s,q}$ are the beamforming vectors corresponding to $p$-th communication user and $q$-th sensing target, respectively. These beamforming vectors are defined  as follows 
	\begin{align}
		\mathbf{w}_{c,{p}}=\sqrt{P_{c}}\frac{\tilde{\mathbf{h}}_{c,p}}{|| \tilde{\mathbf{h}}_{c,p}|| }, \hspace{0.5cm}\mathbf{w}_{s,q}=\sqrt{P_{s}}\frac{\tilde{\mathbf{h}}_{s,q}}{|| \tilde{\mathbf{h}}_{s,q}|| }
		\label{powdist}
	\end{align}
	where $\tilde{\mathbf{h}}_{c,p}$ is the overall channel for the $p$-th communication user  and {$\tilde{\mathbf{h}}_{s,q}$ is the overall channel for $q$-th sensing target for $p\in\left\lbrace 1,\cdots,K\right\rbrace $ and $q\in\left\lbrace 1,\cdots,M\right\rbrace $. Here, for a communication user on Side-$T$,  $\tilde{\mathbf{h}}_{c,p}=\mathbf{h}_{p,T}\mathbf{\Phi}_t\mathbf{G}$ (\ref{received-kt}),  while for those on Side-$R$, it is $\tilde{\mathbf{h}}_{c,p}=\mathbf{h}_{p,R}\mathbf{\Phi}_r\mathbf{G}+\mathbf{g}_{p,R}$ (\ref{received-kr}).  Similarly,   for a sensing target on Side-$T$ (\ref{targetB}),  } $\tilde{\mathbf{h}}_{s,q}=\mathbf{b}(\varphi_{s,h}^{q,T},\varphi_{s,v}^{q,T})\mathbf{\Phi}_t\mathbf{G}$, while for ones on Side-$R$ (\ref{received-kr}), it is $\tilde{\mathbf{h}}_{s,q}=\mathbf{a}(\theta^{{q,R}}_{s,h},\theta^{{q,R}}_{s,v})+\tilde{\mathbf{b}}(\varphi^{{q,R}}_{s,h} ,\varphi^{{q,R}}_{s,v})\mathbf{\Phi}_r\mathbf{G}$. Additionally, the transmit power for each communication user and sensing target is represented as $P_c$ and $P_s$, respectively, \mbox{where $P_c=P_s=P_{\text{BS}}/(K+M)$}.
	\subsection{Cramer-Rao Bound (CRB) Analysis}
	In this section, to evaluate sensing performance of the targets, the CRB estimation for  2D AoDs of the targets has been investigated. To achieve this, first, the Fisher information matrix (FIM) is derived. 
	
	Let $\boldsymbol{\zeta}_{m,i}\in\mathbb{C}^{4\times1}=[\boldsymbol{\varphi}^{m,i}, {\boldsymbol{\beta}}^{m,i}]^{\mathrm{T}}$ be the vector of unknown parameters to be estimated that includes 2D AoDs of $m$th target  on Side-$i$, ${\boldsymbol{\varphi}^{m,i}}\in\mathbb{C}^{1\times2}$ and its corresponding complex channel coefficient $\boldsymbol{\beta}^{m,i}\in\mathbb{C}^{1\times2}$, where $i\in\mathcal{S}=\left\lbrace T,R\right\rbrace $ and $m\in\left\lbrace1,\cdots, M_i \right\rbrace $. For the {} target on the  Side-$T$,  ${\boldsymbol{\varphi}^{m,T}}=[ {\varphi}^{m,T}_{s,h}, {\varphi}^{m,T}_{s,v} ]$ and  complex channel coefficient $\boldsymbol{\beta}^{m,T}=[\mathfrak{Re}({\beta}^{m,T}_{\text{RIS}}), \mathfrak{Im}({\beta}^{m,T}_{\text{RIS}})]$. Then, in order to obtain the FIM,  overall received echo signal in (\ref{receivedt}) is vectorized for $\mathbf{u}_{m,T}=\beta_{\text{RIS}}^{m,T}\vect(\mathbf{G}^{\mathrm{H}}\mathbf{\Phi}^{\mathrm{H}}_t\mathbf{B}_m\mathbf{\Phi}_t\mathbf{G}\mathbf{X})$, $\tilde{\mathbf{n}}_{m,T}=\vect({\tilde{\mathbf{N}}_s^{m,T}})$ as follows
	\begin{equation}
		\mathbf{y}_s^{m,T} = \mathbf{u}_{m,T}+\tilde{\mathbf{n}}_{m,T}. 
	\end{equation}
	Therefore, the FIM for estimating $\boldsymbol{\zeta}_{m,T}$ is constructed as \cite{bekkerman2006target}
	\begin{equation}
		\mathbf{J}_{{m,T}}=\begin{bmatrix} \mathbf{J}_{\boldsymbol{\varphi}\boldsymbol{\varphi}}^{m,T} & \mathbf{J}_{\boldsymbol{\varphi{{\beta}}}}^{m,T} \\ \mathbf{J}_{\boldsymbol{{{\beta}}\varphi}}^{m,T} & \mathbf{J}_{\boldsymbol{{\beta}{{\beta}}}}^{m,T}
		\end{bmatrix}.
		\label{fim}
	\end{equation}
	Then,  utilizing (\ref{fim}), the CRB can be expressed as \cite{bekkerman2006target}
	\begin{equation}
		\crb(\boldsymbol{\varphi}_{m,T})= [\mathbf{J}_{\boldsymbol{\varphi\varphi}}^{m,T}-\mathbf{J}_{\boldsymbol{\varphi{\beta}}}^{m,T} ({\mathbf{J}_{\boldsymbol{\beta\beta}}}^{m,T})^{-1}\mathbf{J}_{{\boldsymbol{\beta\varphi}}}^{m,T} ]^{-1}
		\label{crb}
	\end{equation}
	where \begin{equation}
		\mathbf{J}_{m,T}({\delta,\omega})=\frac{2}{{{\sigma_{\tilde{n}_s}^2}}}\mathfrak{Re}\left\lbrace \frac{\partial\mathbf{u}^{\mathrm{H}}_{m,T}}{\partial\boldsymbol{\zeta}_\delta}\frac{\partial\mathbf{u}_{m,T}}{\partial\boldsymbol{\zeta}_\omega}\right\rbrace\hspace{0.2cm}{\delta},{\omega}\in\left\lbrace 1,2,3,4\right\rbrace 
	\end{equation}
	Here, $\mathbf{B}_m=\mathbf{b}^{\mathrm{H}}(\varphi_{s,h}^{m,T},\varphi_{s,v}^{m,T})\mathbf{b}(\varphi_{s,h}^{m,T},\varphi_{s,v}^{m,T})$  
	{\begin{align}
			&\frac{\partial\mathbf{u}_{m,T}}{\partial\boldsymbol{\varphi}_{m,T}}=\nonumber\\
			&\big[\beta_{\text{RIS}}^{m,T}\vect(\mathbf{G}^{\mathrm{H}}\mathbf{\Phi}^{\mathrm{H}}_t\dot{\mathbf{B}}_{{\varphi}_h}^m\mathbf{\Phi}_t\mathbf{G}\mathbf{X}), 
			\beta_{\text{RIS}}^{m,T}\vect(\mathbf{G}^{\mathrm{H}}\mathbf{\Phi}^{\mathrm{H}}_t\dot{\mathbf{B}}_{{\varphi}_{v}}^m\mathbf{\Phi}_t\mathbf{G}\mathbf{X})\big]\\
			&\frac{\partial\mathbf{u}_{m,T}}{\partial\boldsymbol{\beta}_{m,T}}= \vect(\mathbf{G}^{\mathrm{H}}\mathbf{\Phi}^{\mathrm{H}}_t\mathbf{B}_m\mathbf{\Phi}_t\mathbf{G}\mathbf{X}) [1 \hspace{0.2cm}j]
	\end{align}}where $\dot{\mathbf{B}}_{{\varphi}_{h}}^m=\frac{\partial\mathbf{B}_m}{\partial{\varphi}_{s,h}^{m,T}}$ and $\dot{\mathbf{B}}_{{\varphi}_v}^m=$ {$\frac{\partial\mathbf{B}_m}{\partial{\varphi}_{s,v}^{m,T}}$.
		Therefore,  since $\mathbf{b}(\varphi_{s,h}^{m,T},\varphi_{s,v}^{m,T})$} $=e^{j\eta_{\text{RIS}}\left( \mathbf{k}_x\sin(\varphi^{m,T}_{s,h})\cos(\varphi^{m,T}_{s,v})+\mathbf{k}_y\sin(\varphi^{m,T}_{s,h})\sin(\varphi^{m,T}_{s,v})\right) }$  as expressed in (\ref{bb}), where  $\mathbf{k}_x$ and $\mathbf{k}_y$  are vectors with scalar elements,  $\dot{\mathbf{B}}_{{\varphi}_{h}}^m$ and $\dot{\mathbf{B}}_{{\varphi}_{v}}^m$ can be calculated as
	{\small\begin{align}
			&	\scalemath{0.95}{\dot{\mathbf{B}}_{{\varphi}_h}}={j\eta_{\text{RIS}}\big( \mathbf{k}_x^{\mathrm{H}}\cos(\varphi^{m,T}_{s,h})\cos(\varphi^{m,T}_{s,v})\hspace{-0.05cm}+\hspace{-0.05cm}\mathbf{k}_y^{\mathrm{H}}\cos(\varphi^{m,T}_{s,h})\sin(\varphi^{m,T}_{s,v}\hspace{-0.04cm})\big)}\nonumber\\
			&\scalemath{0.9}{\times \big(-\mathbf{b}(\varphi_{s,h}^{m,T},\varphi_{s,v}^{m,T})^{\mathrm{H}}\mathbf{b}(\varphi_{s,h}^{m,T},\varphi_{s,v}^{m,T})+\mathbf{b}(\varphi_{s,h}^{m,T},\varphi_{s,v}^{m,T})\mathbf{b}(\varphi_{s,h}^{m,T},\varphi_{s,v}^{m,T})^{\mathrm{H}}\big)}\label{12}\\
			&	\scalemath{0.95}{\dot{\mathbf{B}}_{{\varphi}_v}=j\eta_{\text{RIS}}\big(\hspace{-0.09cm}-\hspace{-0.06cm} \mathbf{k}_x^{\mathrm{H}}\sin(\varphi^{m,T}_{s,h})\sin(\varphi^{m,T}_{s,v})\hspace{-0.05cm}+\hspace{-0.05cm}\mathbf{k}_y^{\mathrm{H}}\sin(\varphi^{m,T}_{s,h})\cos(\varphi^{m,T}_{s,v})\big)}\nonumber\\
			&\scalemath{0.9}{\times \big(-\mathbf{b}(\varphi_{s,h}^{m,T},\varphi_{s,v}^{m,T})^{\mathrm{H}}\mathbf{b}(\varphi_{s,h}^{m,T},\varphi_{s,v}^{m,T})+\mathbf{b}(\varphi_{s,h}^{m,T},\varphi_{s,v}^{m,T})\mathbf{b}(\varphi_{s,h}^{m,T},\varphi_{s,v}^{m,T})^{\mathrm{H}}\big)}\label{13}
	\end{align}}Let the overall signal be represented as $\mathbf{F}_m=\mathbf{G}^{\mathrm{H}}\mathbf{\Phi}^{\mathrm{H}}_t{\mathbf{B}}_m\mathbf{\Phi}_t\mathbf{G}\mathbf{X}$, while  $\dot{\mathbf{F}}_{\varphi_h}=\mathbf{G}^{\mathrm{H}}\mathbf{\Phi}^{\mathrm{H}}_t\dot{\mathbf{B}}_{{\varphi}_h}^m\mathbf{\Phi}_t\mathbf{G}\mathbf{X}$ and $\dot{\mathbf{F}}_{\varphi_v}=\mathbf{G}^{\mathrm{H}}\mathbf{\Phi}^{\mathrm{H}}_t\dot{\mathbf{B}}_{{\varphi}_v}^m\mathbf{\Phi}_t\mathbf{G}\mathbf{X}$. Therefore,
	the elements of the FIM can be  defined as \cite{song2023intelligent}
	{\begin{align}
			&\mathbf{J}_{{\boldsymbol{\varphi\varphi}}}^{m,T}(\mathbf{F}_m)= \nonumber\\
			&\frac{2L}{{\sigma_{\tilde{n}_s}^2}}(\beta_{\text{RIS}}^{m,T})^2\mathfrak{Re}\left\lbrace\begin{bmatrix}
				\Tr(\mathbf{\dot{F}}_{\varphi_h}\mathbf{R_x}\dot{\mathbf{F}}_{\varphi_h}^\mathrm{H}) & \Tr(\mathbf{\dot{F}}_{\varphi_h}\mathbf{R_x}\dot{\mathbf{F}}_{\varphi_v}^\mathrm{H}) \\ \Tr(\mathbf{\dot{F}}_{\varphi_v}\mathbf{R_x}\dot{\mathbf{F}}_{\varphi_h}^\mathrm{H}) & \Tr(\mathbf{\dot{F}}_{\varphi_v}\mathbf{R_x}\dot{\mathbf{F}}_{\varphi_v}^\mathrm{H}) 
			\end{bmatrix} \right\rbrace \label{fim1}\\
			& \mathbf{J}_{\boldsymbol{\varphi}{\boldsymbol{\beta}}}^{m,T}(\mathbf{F}_m)=\frac{2L}{{\sigma_{\tilde{n}_s}^2}}\mathfrak{Re}\left\lbrace\begin{bmatrix}
				\beta_{\text{RIS}}^{m,T}\Tr(\mathbf{{F}}_{m}\mathbf{R_x}\dot{\mathbf{F}}_{\varphi_h}^\mathrm{H}) \\ 	\beta_{\text{RIS}}^{m,T}\Tr(\mathbf{{F}}_{m}\mathbf{R_x}\dot{\mathbf{F}}_{\varphi_v}^\mathrm{H}) 
			\end{bmatrix}\begin{bmatrix}
				1&j
			\end{bmatrix} \right\rbrace\\
			&\mathbf{J}_{\boldsymbol{{\beta\beta}}}^{m,T}(\mathbf{F}_m)=\frac{2L}{{\sigma_{\tilde{n}_s}^2}}\mathbf{I}_2\Tr(\mathbf{F}_m\mathbf{R_x}\mathbf{F}_m^\mathrm{H})
			\label{crbf}
	\end{align}}Then, the CRB of the $m$th {distant} target is obtained by substituting  (\ref{fim1}-\ref{crbf}) into (\ref{crb}).
	
	In similar way, the CRB of  the sensing targets on Side-$R$ can be calculated for AoDs from BS to the target O$_{m,R}$ ${\boldsymbol{\theta}^{m,R}}=[ {\theta}^{m,R}_{s,h}, {\theta}^{m,R}_{s,v} ]$ and  complex channel coefficient $\boldsymbol{\beta}^{m,R}=[\mathfrak{Re}({\beta}^{m,R}_{\text{BS}}), \mathfrak{Im}({\beta}^{m,R}_{\text{BS}})]$ via following (\ref{crb}-\ref{crbf}) steps.

	

	\begin{table}[!t]
			\vspace{-0.2cm}
		\caption{System Parameters}
		\centering
		\setlength{\tabcolsep}{2.4pt} 
		{\scriptsize	\begin{tabular}{|c||l|c|}
				\hline
				\textbf{Notation} & \textbf{Parameter} & \textbf{Value}\\ \hline \hline
				$f_c$&  Operating frequency &  $3.315$ GHz   \\ \hline
				$T_x$&  Number of transmit/receive antennas at BS &  $8$   \\ \hline
				$K$		& Total number of communication users & $2$  \\ \hline
				$M$	&Total number of sensing targets  &  $2$    \\ \hline
				$\kappa$ & Rician factor& $3$ dB  \\ \hline
				$\sigma_{n_{s}}^2=\sigma_{n_{c}}^2$& Noise power& $-80$ dBm \\ \hline
				$L_c$  & Coherence time length & 100 \\ \hline
				$\alpha_0$  & Reference path loss at $1$ m & $30$ dB\\ \hline
				$\rho$  & Path loss exponent & $2.2$ \\ \hline
				$\Lambda$ & RCS & $100$ $\text{m}^2$ \\ \hline
				$\theta_h^R, \theta_v^R$  & AoDs of target O$_R$ & $\left\lbrace35^\circ, 110^\circ \right\rbrace $  \\ \hline
				$\varphi_h^T, \varphi_v^T$  & AoDs of target O$_T$ & $\left\lbrace 40^\circ, 108^\circ\right\rbrace  $     \\ \hline
				$d_{\text{BS-RIS}}$&  Distance: BS-STAR-RIS  & $5$ m \\ \hline
				$d_{\text{RIS-O}_T}$, $d_{\text{RIS-U}_T}$&  Distances: STAR-RIS-O$_T$  and STAR-RIS-U$_T$& $\left\lbrace 17, 18 \right\rbrace $ m \\ \hline
				$d_{\text{BS-O}_R}$, $d_{\text{RIS-O}_R}$ &  Distances: BS-O$_R$   and STAR-RIS-O$_R$  & $\left\lbrace 38,41 \right\rbrace $ m  \\ \hline
				{$d_{\text{BS-U}_R}$, $d_{\text{RIS-U}_R}$} &  Distances: BS-U$_R$   and STAR-RIS-U$_R$  & $\left\lbrace 25, 27 \right\rbrace $ m  \\ \hline
		\end{tabular}}
		\label{systable}
	\end{table}

	\section{Simulation Results}
	In this section, Monte Carlo simulation results are provided to evaluate the sensing performance of the  hybrid STAR-RIS-assisted ISAC scheme  and the proposed optimization algorithm. {Computer} simulation results are conducted for a two-user and two-target scenario, where there is one user and one target located on each side ($T,R$) of the hybrid STAR-RIS. Specifically, the user and target at Side-$R$  donated as U$_R$ and O$_R$, while those on Side-$T$ are referred to as U$_T$ and O$_T$, respectively. Additionally, system parameters used in the simulations are listed in Table \ref{systable}.
	
			\begin{figure}[t]
		\centering
		\includegraphics[width=1\columnwidth]{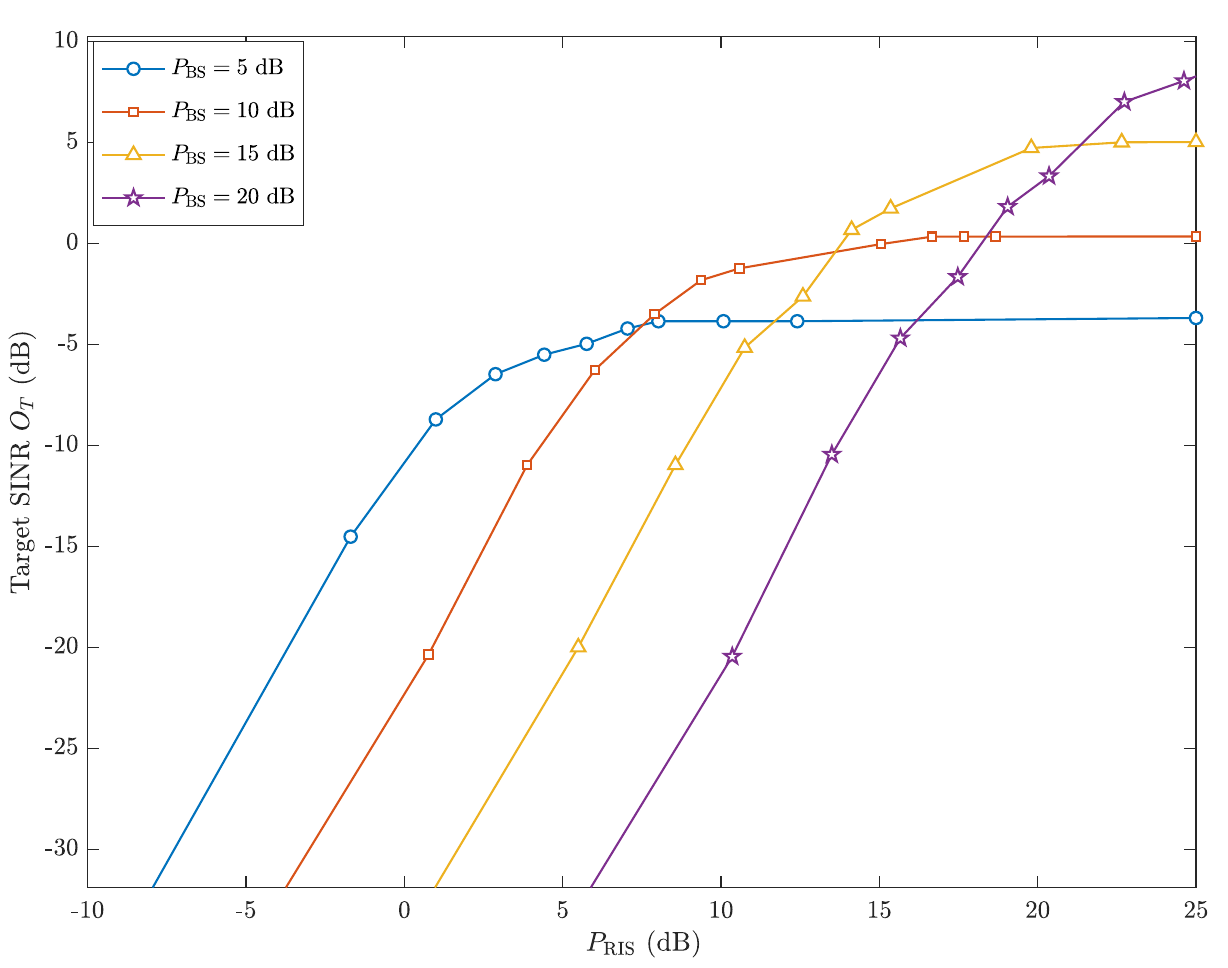}	
		\caption{SINR of  target O$_T$ versus power consumption at active STAR-RIS elements for $N=36$ and different $P_{\text{BS}}$ values. }
		\label{fig.sinrb}
	\end{figure}
	\begin{figure}[t]
	\centering
	\subfloat[]{%
		\includegraphics[width=0.492\linewidth]{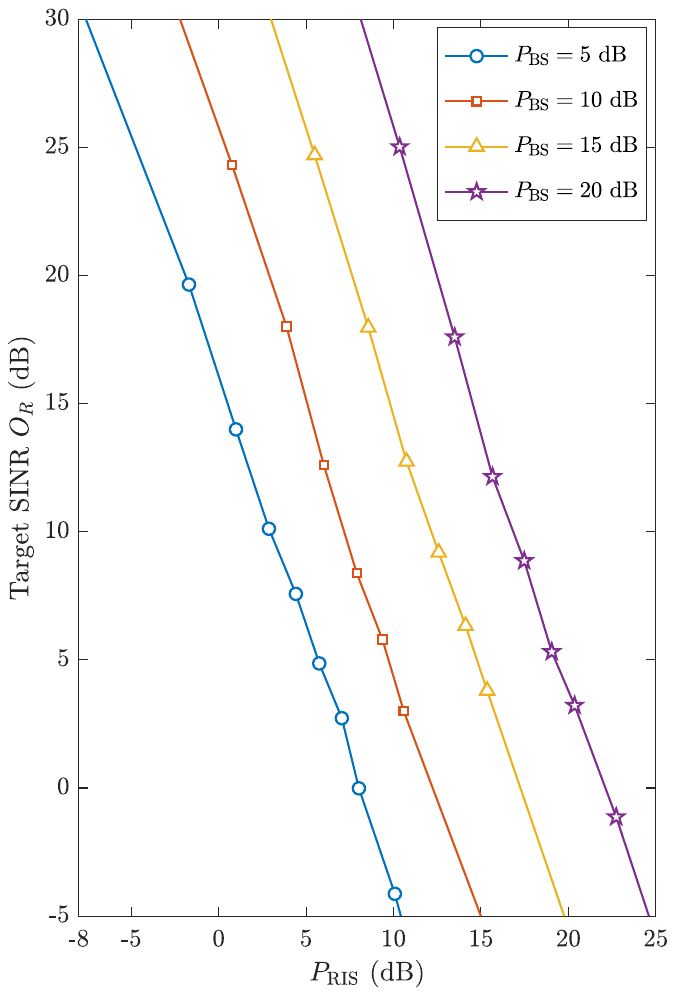}}
	\label{4a}
	\centering
	\subfloat[]{%
		\includegraphics[width=0.485\linewidth]{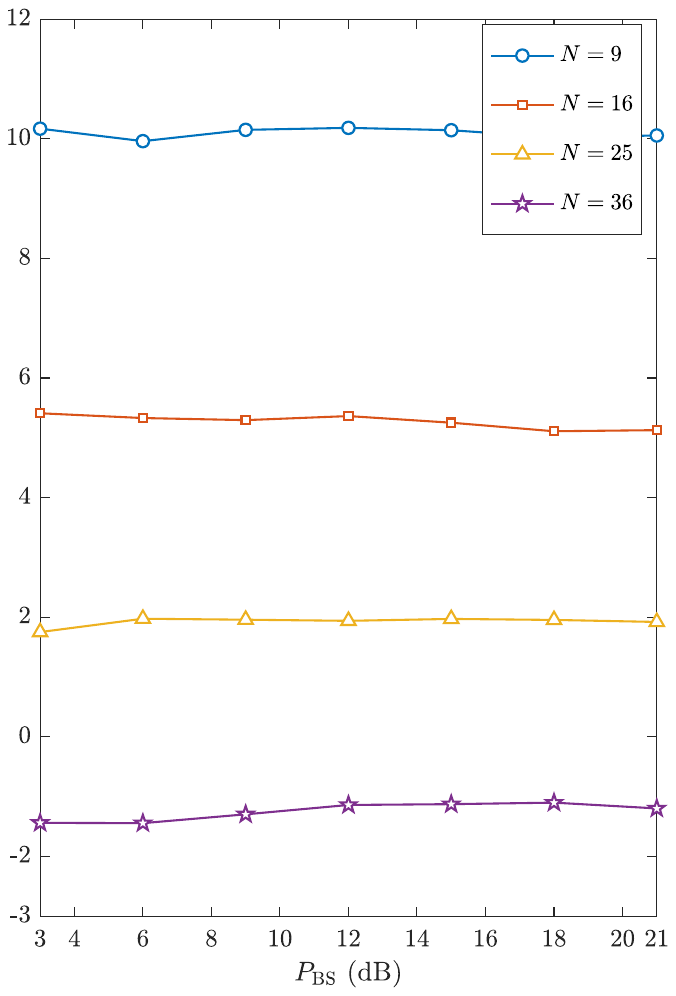}}
	\label{4b}
	\caption{ SINR of target $O_T$ for varying $P_\mathrm{RIS}$ (a) and {$P_\mathrm{BS}$ (b).}}	\label{fig.sinra}
\end{figure}
	
	Fig. \ref{fig.sinrb} presents the SINR of the target on Side-$T$, denoted as O$_T$, for $N=36$ hybrid STAR-RIS elements with varying $P_{\text{RIS}}$ due to increasing $P_A$
	 and transmit power levels at the BS, $P_{\text{BS}} = \left\lbrace 5, 10, 15, 20\right\rbrace $ dB. The results clearly show that increasing $P_{\text{BS}}$ enhances the SINR performance of the target O$_T$. On the other hand, higher $P_{\text{BS}}$ values also lead to greater $P_{\text{RIS}}$, as discussed in (\ref{pris}). {This highlights a fundamental trade-off between enhancing SINR and increasing the power consumption of the active STAR-RIS elements, where 
$P_{\text{BS}}$
	plays a crucial role in balancing the performance improvements with power consumption.} Moreover, it can be deduced from Fig.  \ref{fig.sinrb}  that after a certain  $P_{\text{RIS}}$ value, the increased thermal noise from the active transmissive elements causes the target SINR values to saturate. This further supports that remaining at lower $P_{\text{RIS}}$ range is significant for maintaining efficient target SINR performance.}
	
In Fig. \ref{fig.sinra}, the SINR of the target on Side-$R$, denoted as O$_R$, is given for varying $P_{\text{RIS}}$  (a) and $P_{\text{BS}}$ (b) values. Fig. 4a), indicates  a degradation in  SINR of the target with increasing $P_{\text{RIS}}$.   {This effect can be attributed to the increase in power of the active elements that enhance the signal for the {distant} target O$_T$, which then serves as interference for O$_R$, as described in (\ref{sinrr}). Consequently, this interference leads to a reduction in the SINR of the target O$_R$. }

{On the other hand,  in Fig. 4b), the SINR of the target $O_R$ given for increasing $P_{\mathrm{BS}}$ values. The results indicate that the increased $P_{\mathrm{BS}}$ is equally distributed among all users and targets in the system, there is a negligible change in target SINR of $O_R$. This can be attributed to the fact that  the increased $P_\mathrm{BS}$  is equally shared among all the users and targets in the system, it does not lead to significant improvements in the SINR of the target $O_R$. Additionally, from Fig. 4b), when $P_{\mathrm{BS}}$ is kept constant, an increase in $N$ degrades the SINR of the target $O_R$. This can be explained by the fact that due to the  direct link between the BS and $O_R$, and an increase in the number of passive elements $N$ has a negligible effect on the SINR of $O_R$, unless the number of elements becomes significantly large. On the other hand, the SINR of the target $O_T$	further improves with increasing number of active transmissive elements, which results in additional interference for $O_R$.}

	\begin{figure}[!t]
	\centering
	{\vspace*{-0.2cm}	\includegraphics[width=1\columnwidth]{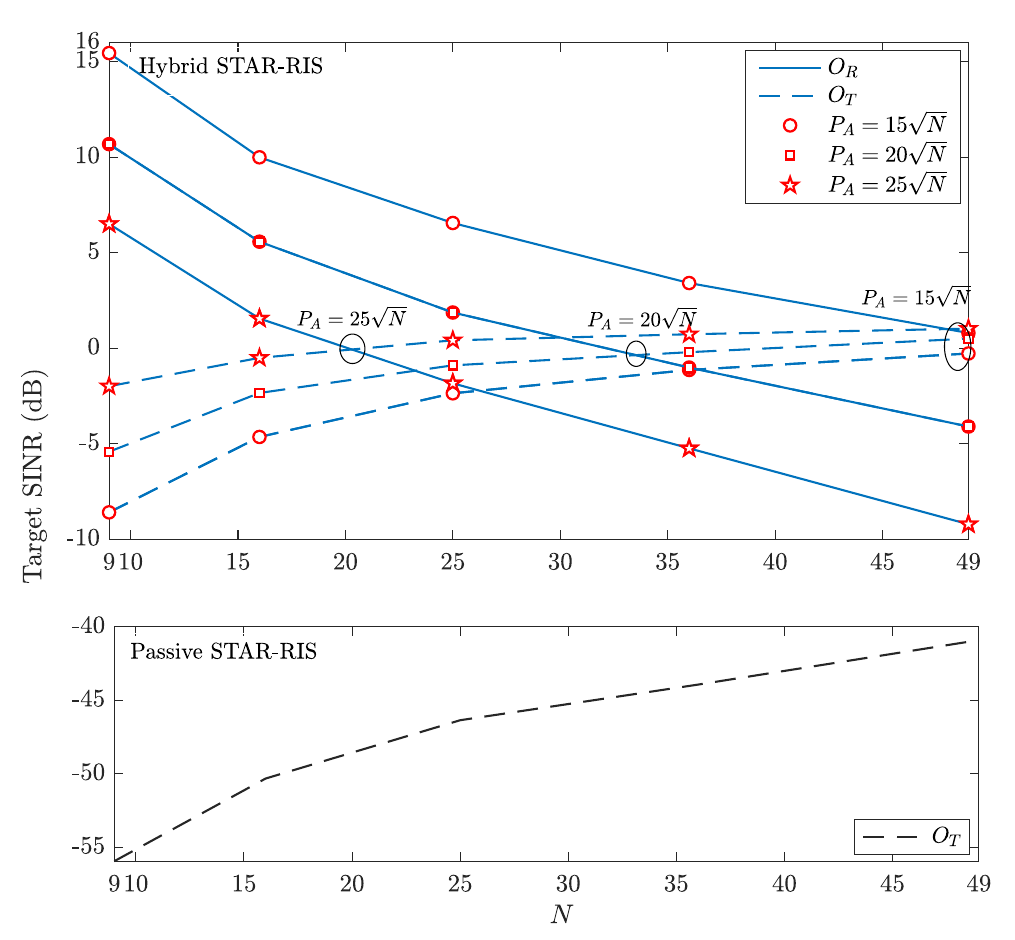}
		\caption{SINR of the targets O$_R$ and O$_T$ for varying numbers of STAR-RIS elements $N$.}
		\label{fig.sinrab}}
\end{figure}

	Fig. \ref{fig.sinrab} presents the target SINR performance for the proposed hybrid STAR-RIS-aided ISAC system {and fully-passive STAR-RIS-aided ISAC systems} for varying number of STAR-RIS elements $N$ at $P_\mathrm{BS}=10$ dB. {First, for a fair comparison, the transmit power of the reference passive RIS system is equated to the total power consumption of the proposed hybrid STAR-RIS assisted system, which includes the additional power consumption introduced by the use of active transmissive elements. The transmit power of the benchmark scheme is assumed to be $P_{\mathrm{Total}}=P_{\mathrm{BS}}+P_{\mathrm{RIS}} $, where the amplification factor is set to  $P_A=25\sqrt{N}$. However, the results indicate that although 
$P_{\mathrm{Total}}$ reaches $16$ dB for $N=49$ the reference passive STAR-RIS-assisted ISAC scheme is unable to mitigate the attenuation of the sensing signal caused by multi-hop transmission.} In contrast, the hybrid STAR-RIS-aided ISAC system outperforms the benchmark, demonstrating a significant improvement in target SINR performance. This improvement can be attributed to the amplifying capability of the transmissive elements in the proposed hybrid STAR-RIS structure, which contrasts with the passive-only reflection/tranmission characteristics  of the benchmark scheme.
		
\begin{figure}[!t]
	\centering
	{	\includegraphics[width=1\columnwidth]{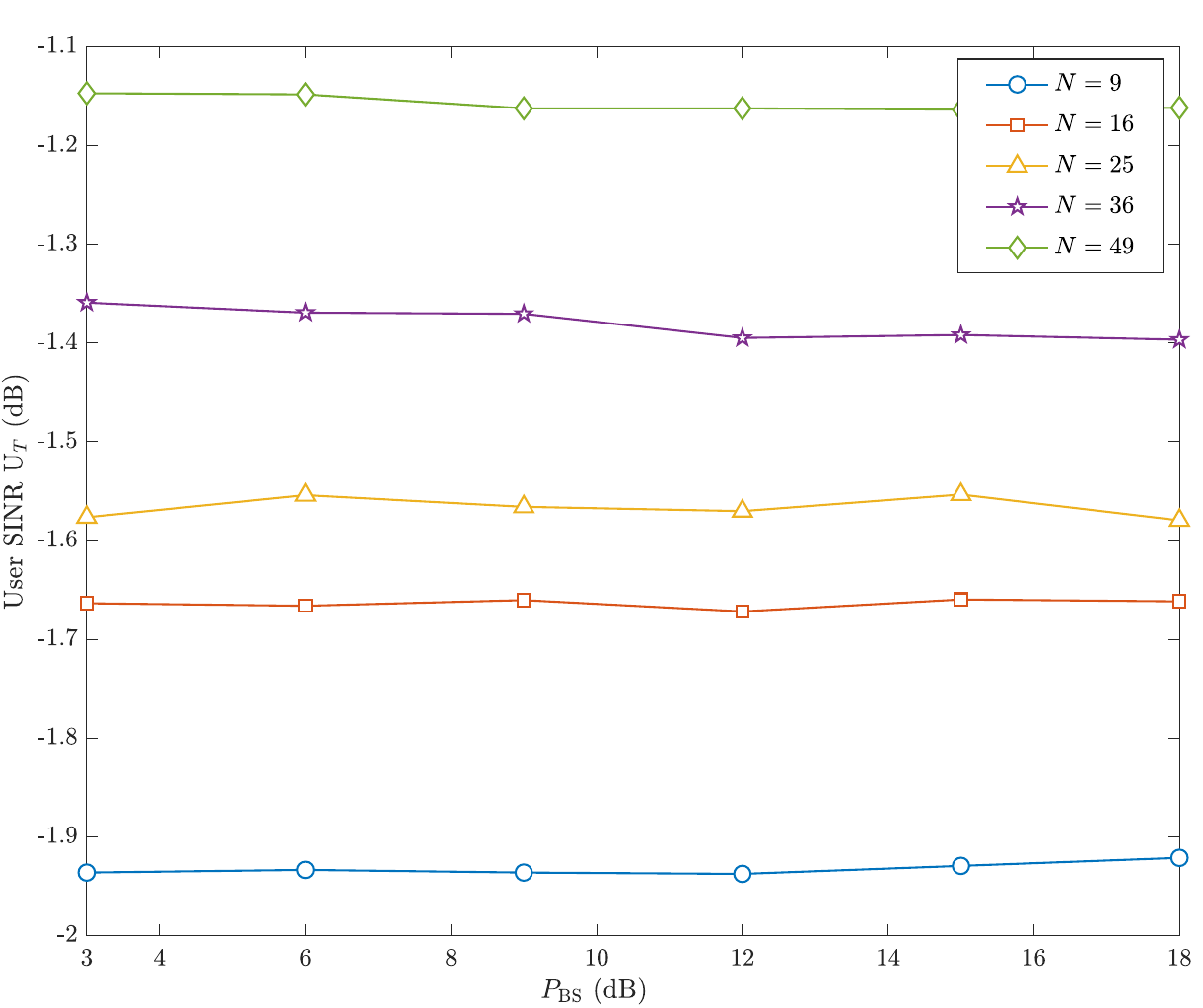}
		\caption{{SINR of the communication user U$_T$  for varying $N$}.} 
		\label{fig.sinru}}
\end{figure}		
		Additionally,  as  illustrated in the   Fig. \ref{fig.sinrab}, for the proposed hybrid STAR-RIS-aided ISAC scheme, increasing $N$ improves the SINR performance of the target O$_T$. However, since the enhanced signal of the target O$_T$ acts as interference for target O$_R$, it simultaneously degrades the SINR of O$_R$, which aligns with the results shown in Figs. \ref{fig.sinrb}-\ref{fig.sinra}. Moreover, the results indicate that increasing the overall amplification factor $P_A$ further enhances the SINR of the target O$_T$ while degrading the SINR performance of the target O$_R$.  These results suggest a fair SINR distribution between the targets on both sides for different $N$ values. Specifically, the SINR of targets O$_R$ and O$_T$ nearly reach equilibrium for varying $P_A$ values. A balanced target SINR is achieved between both sides at approximately  $P_A = 15\sqrt{N}$ around $N = 49$, for $P_A = 20\sqrt{N}$ when $N = 34$, and for $P_A = 25\sqrt{N}$ when $N = 20$. This suggests that for smaller amplification factors $P_A$ a hybrid STAR-RIS with a larger $N$ is needed to maintain a balanced SINR distribution between both sides.

	\begin{figure}[t]
		\centering
		\includegraphics[width=1\columnwidth]{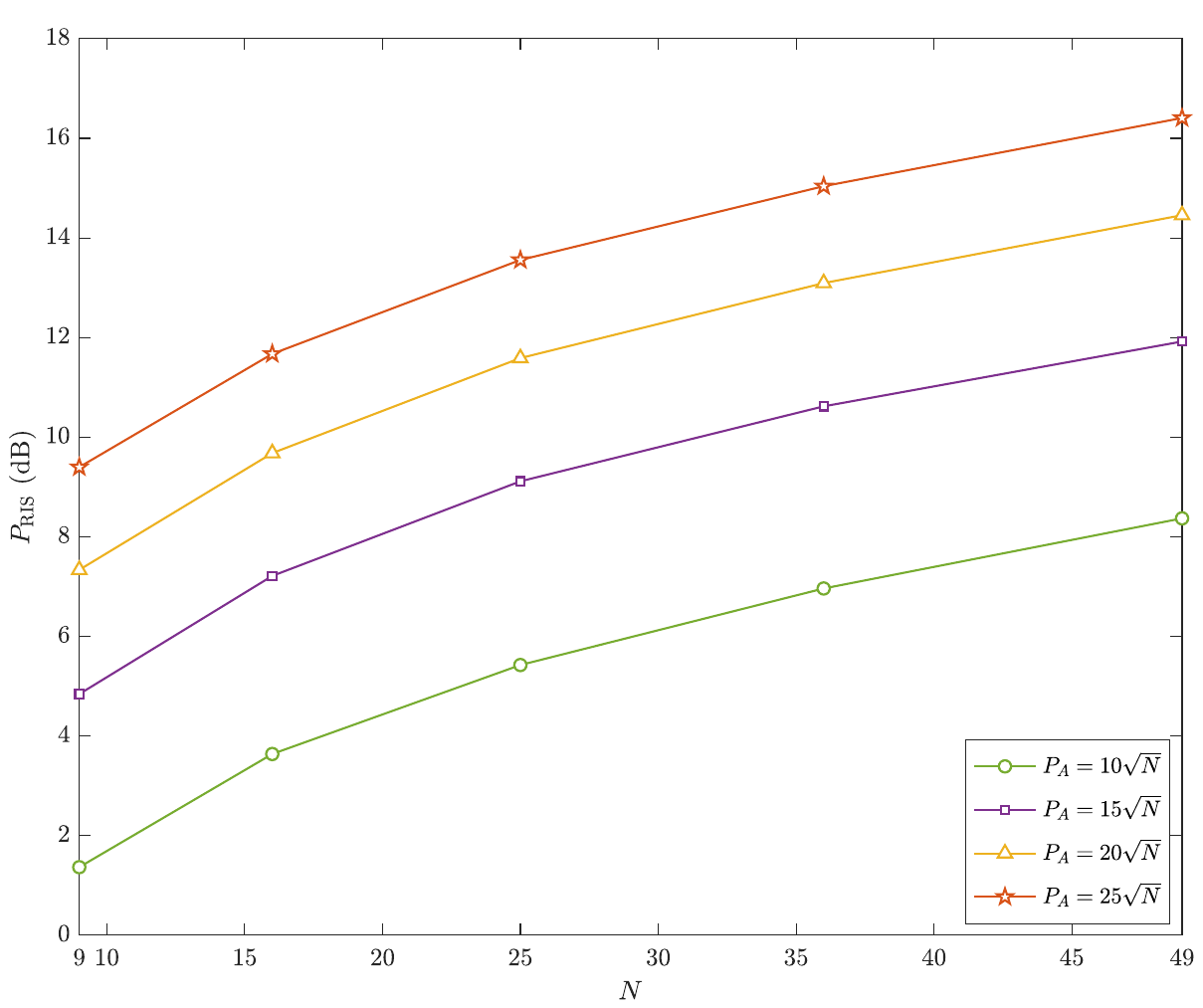}
		\caption{Power consumption of active transmissive elements $P_{\text{RIS}}$ at $P_{\text{BS}}=10$ dB for varying  $N$.} 
		\label{fig.pow}
	\end{figure}

	\begin{figure}[t]
	\centering
	\vspace{0.15cm}
	\includegraphics[width=1.0\columnwidth]{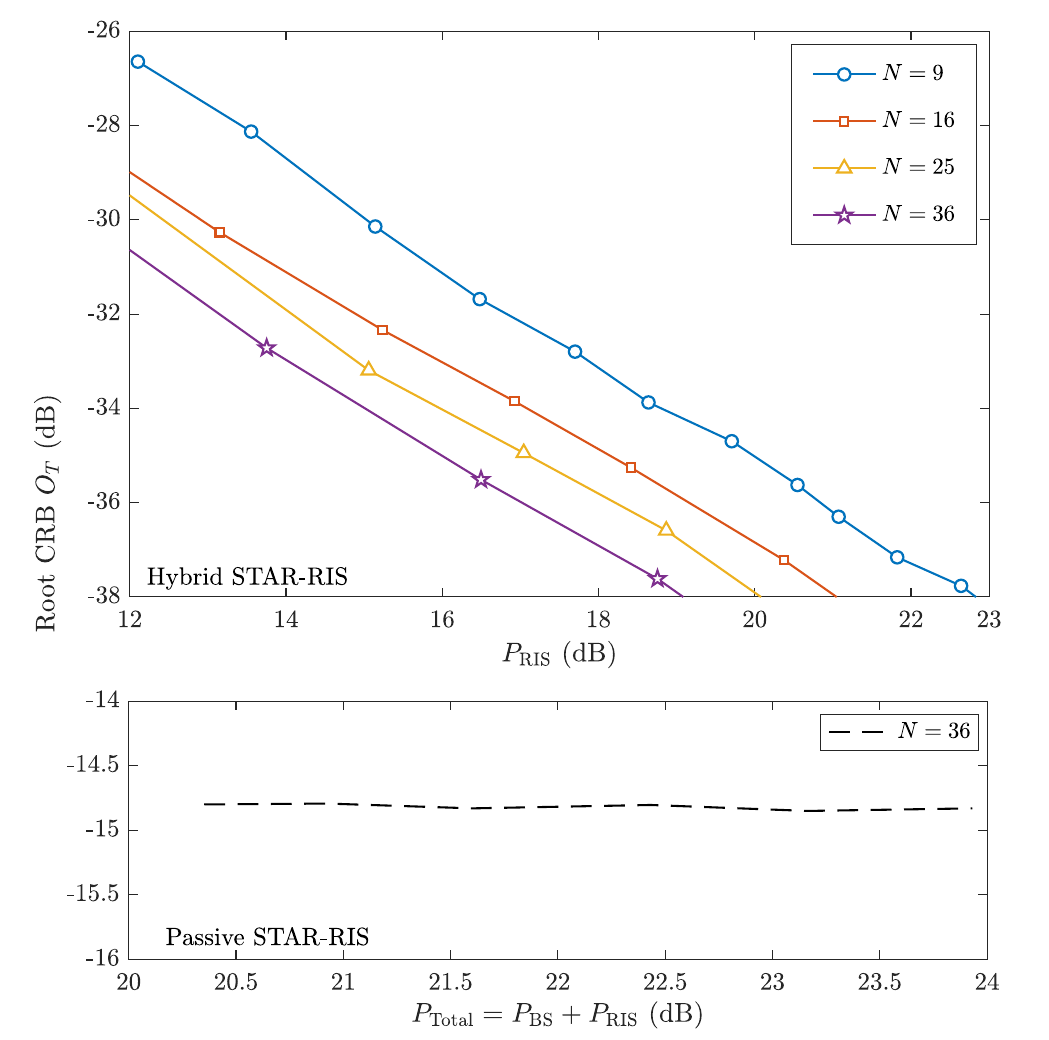}
	\caption{Root CRB of target O$_T$ versus power consumption of active transmissive elements $P_{\text{RIS}}$ for varying numbers of hybrid  STAR-RIS elements $N$. }
	\label{fig.crbb}
\end{figure}

	{In Fig. \ref{fig.sinru}, the SINR performance of the communication user on Side-T, U$_T$  for varying $N$  is shown.  As anticipated, the results indicate  significant  improvement in SINR performance of the user U$_T$ as $N$ increases. On the other hand, an increase in the  $P_{\text{BS}}$ dB does not yield in a noticeable enhancement  in the  SINR performance of the user U$_T$. It can be deduced from Fig. \ref{fig.sinru} that since signal enhancement on Side-$T$ is achieved via active transmissive elements, the transmit power at the BS, as a result, $P_{\text{BS}}$, has an almost negligible effect on the user SINR on Side-$T$.}
	
	In Fig. \ref{fig.pow}, the power consumption of active transmissive elements $P_{\text{RIS}}$ is given for varying numbers of hybrid STAR-RIS elements $N$ and $P_A$. As corroborated by (\ref{pris}) and the results presented  in Figs. \ref{fig.sinrab}-\ref{fig.sinru} is obvious that an increase in $N$ leads to  higher  $P_{\text{RIS}}$. Particularly, for a certain $N$ value, a $5\sqrt{N}$  increase in  $P_{A}$ results in $2$ dB increase in $P_{\text{RIS}}$. {These results illustrate a trade-off between achieving high target SINR and user SINR   while  managing power consumption of the active transmissive elements.}
	
	Fig. \ref{fig.crbb} illustrates the root CRB 
	estimation of  2D AoDs of the target  O$_T${ for the proposed hybrid STAR-RIS and reference passive STAR-RIS-assisted ISAC schemes }  for different $N$ values at $P_{\text{BS}}=20$ dB. Here, {although the benchmark passive STAR-RIS-aided ISAC scheme is evaluated for the overall power $P_\mathrm{Total}=P_\mathrm{BS}+P_\mathrm{RIS}$, it  achieves a negligible improvement in root CRB performance compared to hybrid STAR-RIS-assisted ISAC scheme. This limited performance improvement of the passive STAR-RIS-assisted scheme  is due to the multi-user and multi-target structure, where an increase in $P_\mathrm{Total}$
		equally impacts both sensing targets and communication users (\ref{powdist}), leading to minimal changes in performance}. On the other hand, for the proposed hybrid STAR-RIS-assisted ISAC scheme, as $N$ increases, there is an noticeable improvement in the root CRB estimation. This can be attributed to  enhanced sensing signal power and an accompanying rise in $P_{\text{RIS}}$ as $N$ increases. {Consequently, while increasing overall transmit power impacts sensing targets and communication users equally, increasing the power of active transmissive elements in the hybrid STAR-RIS scheme specifically improves the performance of the sensing target and the communication user on Side-$T$. }

		\begin{figure}[t]
		\centering
		\includegraphics[width=1\columnwidth]{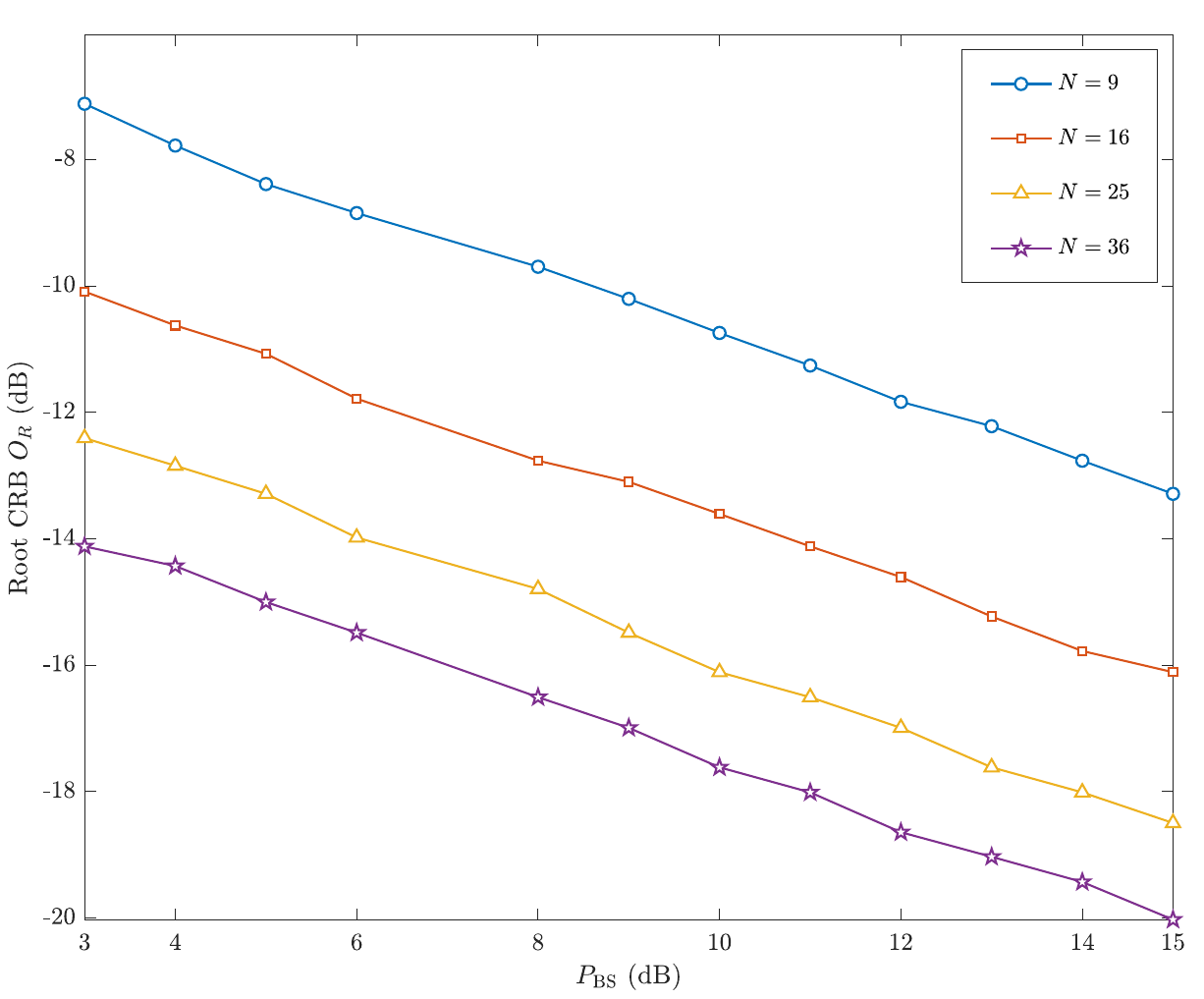}
		\vspace{-0.3cm}
		\caption{Root CRB of target O$_R$ versus  $P_{\text{BS}}$ for varying numbers of hybrid  STAR-RIS elements $N$. }
		\label{fig.crba}
	\end{figure}
	
	In Fig. \ref{fig.crba}, the root CRB estimation for AoDs of the target O$_R$ is given for varying $P_\text{BS}$. It is apparent from the results that an increase in   $P_{\text{BS}}$ and  $N$  improve the CRB estimation of AoDs of the target O$_R$. It is also observed that when root CRB is kept constant, a hybrid STAR-RIS comprising a higher number of $N$ elements necessitate a lower $P_{\text{BS}}$ at the BS.

	\section{Conclusion}
	
	This paper has  proposed a novel hybrid STAR-RIS-aided ISAC transmission scheme with full-space communication and sensing capabilities for  multi-user and multi-target scenarios. In the proposed scheme, to address significant path attenuation resulting from multi-hop transmission, low-power active transmissive elements have been  considered at  the hybrid STAR-RIS that introduce considerable power consumption. Moreover, sensing performance metrics including target SINR and CRB for 2D AoDs estimation have been derived.  An SDR-based optimization algorithm has been developed  to maximize the minimum SINR of the targets by optimizing transmissive and reflective coefficients of the hybrid STAR-RIS elements. Furthermore, to evaluate communication and sensing performance of the system and illustrate the effectiveness of the proposed algorithm, a comprehensive simulations have been conducted.

	\balance
		\bibliographystyle{IEEEtran}
		\bibliographystyle{unsrt} 
		\bibliography{refer_isac,IEEEsettings}
		
		
		

\end{document}